\documentclass{SciPost}

\usepackage{xcolor}
\usepackage{listings}
\usepackage{float}
\usepackage{mathtools}
\usepackage{url}

\hypersetup{
    colorlinks,
    linkcolor={red!50!black},
    citecolor={blue!50!black},
    urlcolor={blue!80!black}
}

\usepackage[bitstream-charter]{mathdesign}
\DeclareSymbolFont{usualmathcal}{OMS}{cmsy}{m}{n}
\DeclareSymbolFontAlphabet{\mathcal}{usualmathcal}

\fancypagestyle{SPstyle}{
\fancyhf{}
\lhead{\colorbox{scipostblue}{\bf \color{white} ~SciPost Physics Codebases }}
\rhead{{\bf \color{scipostdeepblue} ~Submission }}

\fancyfoot[C]{\textbf{\thepage}}
}

\definecolor{codegray}{RGB}{245,245,245}
\definecolor{commentgray}{RGB}{120,120,120}
\definecolor{stringred}{RGB}{200,50,50}
\definecolor{commentgreen}{RGB}{0,180,0}

\lstdefinestyle{shell}{
  language=bash,
  backgroundcolor=\color{codegray},
  basicstyle=\ttfamily\small,
  commentstyle=\color{commentgray},
  stringstyle=\color{stringred},
  keywordstyle=\color{blue},
  showstringspaces=false,
  frame=single,
  rulecolor=\color{black!20},
  breaklines=true,
}

\lstdefinestyle{python}{
  language=Python,
  backgroundcolor=\color{codegray},
  basicstyle=\ttfamily\small,
  keywordstyle=\color{blue},
  commentstyle=\color{commentgreen},
  stringstyle=\color{stringred},
  showstringspaces=false,
  frame=single,
  rulecolor=\color{black!20},
  breaklines=true,
}

\newcommand{\SRC}{svPITE}
\usepackage{xspace}

\begin{document}

\pagestyle{SPstyle}

\begin{center}{\Large \textbf{\color{scipostdeepblue}{
svPITE: A Python package for the state-vector-based probabilistic imaginary-time evolution algorithm
% Article Title, as descriptive as possible, ideally fitting in two lines (approximately 150 characters) or less\\
}}}\end{center}

%%%%%%%%%% AUTHORS
\begin{center}\textbf{
Pascal Sievers \textsuperscript{1,2$\star$} 
and
Satoshi Ejima \textsuperscript{1$\dagger$} 
}\end{center}

\begin{center}
%%%%%%%%%% AFFILIATIONS
{\bf 1} Institute of Software Technology,
        German Aerospace Center (DLR), 
        22529 Hamburg, Germany
% \\

{\bf 2} Department of Physics, 
        University of Hamburg, 
        Notkestraße 9-11,
        22607 Hamburg, Germany
% \\
\\
%%%%%%%%%% EMAIL %%%%%%%%%%
\vspace{\baselineskip}
$\star$ \href{mailto:email1}{\small pascal.sievers@dlr.de}\,,\quad
$\dagger$ \href{mailto:email2}{\small satoshi.ejima@dlr.de}

\end{center}

\section*{\color{scipostdeepblue}{Abstract}}
\textbf{\boldmath{%
We present a Python package for ground-state preparation based on the probabilistic imaginary-time evolution algorithm, with particular focus on its state-vector-based implementation. A standard shot-based simulation is also supported, and results can be benchmarked against exact diagonalisation via a dedicated wrapper. The package enables efficient tuning of initial parameters, facilitating systematic exploration and optimisation of the method’s performance. Starting from the prepared ground state, the strong interoperability with other packages further enables real-time evolution and the computation of spectral functions, such as the spin-spin dynamical structure factor.
% The abstract is in boldface, and should fit in 8 lines. It should be written in a clear and accessible style, emphasizing the context, the problem(s) studied, the methods used, the results obtained, the conclusions reached, and the outlook. You can add a table contents, recommended if your paper is more than 6 pages long.
}}

\vspace{\baselineskip}

%%%%%%%%%% BLOCK: Copyright information
% This block will be filled during the proof stage, and finilized just before publication.
% It exists here only as a placeholder, and should not be modified by authors.
\noindent\textcolor{white!90!black}{%
\fbox{\parbox{0.975\linewidth}{%
\textcolor{white!40!black}{\begin{tabular}{lr}%
  \begin{minipage}{0.6\textwidth}%
    {\small Copyright attribution to authors. \newline
    This work is a submission to SciPost Physics Codebases. \newline
    License information to appear upon publication. \newline
    Publication information to appear upon publication.}
  \end{minipage} & \begin{minipage}{0.4\textwidth}
    {\small Received Date \newline Accepted Date \newline Published Date}%
  \end{minipage}
\end{tabular}}
}}
}
%%%%%%%%%% BLOCK: Copyright information

\vspace{10pt}
\noindent\rule{\textwidth}{1pt}
\tableofcontents
\noindent\rule{\textwidth}{1pt}
\vspace{10pt}

%%%%%%%%% CONTENTS 

\section{Introduction}
\label{sec:intro}

Imaginary-time evolution (ITE) is a central technique in quantum many-body physics, widely used for ground-state preparation and thermal state simulations~\cite{Goldberg1967,RevModPhys.73.33,Schollwoeck2011}.
In recent years, there has been growing interest in implementing imaginary-time dynamics on quantum devices, where non-unitary evolution must be approximated using variational, quantum or probabilistic  approaches~\cite{mcardle2019,Yuan2019,Motta2019,lin2021,kosugi2022,nishi_optimal_2023}.

Probabilistic ITE (PITE) provides one such framework, in which non-unitary evolution $e^{-\hat{\cal H}\Delta\tau}$ is realized via stochastic application of unitary operations combined with measurement and post-selection~\cite{lin2021,kosugi2022,nishi_optimal_2023}. This approach enables the emulation of imaginary-time dynamics while remaining compatible with the constraints of quantum circuits.
While most existing implementations of PITE focus on shot-based realizations that explicitly incorporate sampling and measurement noise, classical simulations based on state-vector representations remain essential for benchmarking and methodological development, as demonstrated in Ref.~\cite{ejima_probabilistic_2025}. In particular, state-vector implementations allow one to isolate intrinsic algorithmic properties such as convergence behaviour and approximation accuracy, independently of statistical fluctuations.

The present codebase provides a reference implementation of PITE using a state-vector backend, while also supporting shot-based simulations within the same framework. It also enables direct comparison with exact diagonalization (ED) results in the same environment, where ED calculations are performed using QuSpin~\cite{weinberg2017}. It is designed for generic quantum spin systems with up to two-body interactions, allowing users to specify a wide class of Hamiltonians in a flexible manner.
The implementation is built on top of Qiskit~\cite{qiskit}, enabling direct use of its circuit abstraction, simulation tools, and built-in parallelization capabilities. This facilitates efficient execution as well as straightforward extension to hardware-oriented workflows.
Compared to existing approaches, the code emphasizes a unified treatment of state-vector and shot-based PITE, providing a transparent environment for cross-validation between idealized and sampling-based dynamics. The modular structure further allows users to explore algorithmic variants and extensions with minimal overhead.
The code reproduces the results presented in Ref.~\cite{ejima_probabilistic_2025}, and most of the figures can be readily regenerated using the provided scripts, ensuring full reproducibility. While Ref.\cite{ejima_probabilistic_2025} and the present work share authors, \SRC{} is a separate implementation, developed with the earlier research code as a reference and reimplemented as a general-purpose package supporting generic Hamiltonians and both state-vector and shot-based PITE simulations.

This paper is organized as follows. Section 2 briefly reviews the theoretical framework of PITE. Section 3 gives an overview of the package. Section 4 describes the basic workflow for using \SRC{}. Section 5 presents examples and benchmarks, and Section 6 outlines future perspectives.

\section{Overview of the PITE algorithm and its state-vector formulation}
\label{sec:PITE}

In this section, we briefly review the PITE algorithm as introduced in Refs.~\cite{kosugi2022,nishi_optimal_2023} and then outline the state-vector formulation derived in Ref.~\cite{ejima_probabilistic_2025}, which forms the basis of the state-vector-based implementation. Throughout this work, we use units in which $\hbar=1$.

\subsection{PITE algorithm}
The following overview of the approximate PITE algorithm closely follows the presentation of Ref.~\cite{ejima_probabilistic_2025}. For an initial state $|\psi_{\rm ini}\rangle$, ITE under Hamiltonian $\hat{\cal H}$ is formally given by
\begin{align}
    |\psi(\Delta\tau)\rangle = 
    \frac{e^{-\hat{\cal H}\Delta\tau}|\psi_{\rm ini}\rangle}{\|e^{-\hat{\cal H}\Delta\tau}|\psi_{\rm ini}\rangle\|},
\end{align}
which is non-unitary and thus not directly implementable in a unitary quantum circuit. To circumvent this, we embed the non-unitary operator 
$\hat{\cal T}\equiv\gamma e^{-\hat{\cal H}\Delta\tau}$ into a larger unitary that acts on the $L$‑qubit system together with a single ancillary qubit. In the ancilla basis
$\{|0\rangle,|1\rangle\}$ the unitary has the block form
\begin{equation}
    \hat{\mathcal{U}}_{\hat{\mathcal{T}}} \equiv \left(\begin{array}{cc}
    \hat{\mathcal{T}} & \sqrt{1-\hat{\mathcal{T}}^2} \\
    \sqrt{1-\hat{\mathcal{T}}^2} & -\hat{\mathcal{T}}
    \end{array}\right)\,.
\end{equation}
Here, the parameter $\gamma$ is initialized at the beginning of the simulation and controls the trade-off between convergence towards the ground state and statistical efficiency of the probabilistic updates.
Determining an optimal or near-optimal value of $\gamma$ is therefore one of the central objectives of \SRC{}. 
To describe the action of the unitary, let the ancilla be initialized in $|0\rangle$, and let $|\psi\rangle$ be an arbitrary state of the $L$-qubit system. 
The unitary $\hat{\mathcal{U}}_{\hat{\mathcal{T}}}$ then acts as
\begin{equation}
    \hat{\mathcal{U}}_{\hat{\mathcal{T}}}|\psi\rangle \otimes|0\rangle
    =\hat{\mathcal{T}}|\psi\rangle \otimes|0\rangle
    +\sqrt{1-\hat{\mathcal{T}}^2}|\psi\rangle \otimes|1\rangle
\end{equation}
thus implementing the non-unitary map $\hat{\mathcal{T}}$ probabilistically. Measuring the ancilla qubit in the $|0\rangle$ state, which occurs with probability $\mathbb{P}_0=\langle\psi| \hat{\mathcal{T}}^2|\psi\rangle$, produces the desired state, proportional to $\hat{\mathcal{T}} |\psi\rangle$.
The complementary outcome applies the Kraus operator $\sqrt{1-\hat{\mathcal{T}}^2}$ and the resulting state is discarded.

\begin{figure}[tb]
    \centering
    \includegraphics[width=0.65\linewidth]{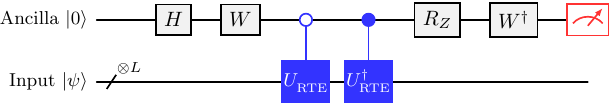}
    \caption{Quantum circuit of the approximate PITE algorithm for a single imaginary-time step, adapted from Ref.~\cite{ejima_probabilistic_2025}. $H$ denotes the Hadamard gate and $R_Z \equiv R_Z(-2\theta_0)$ represents a single-qubit rotation about the $Z$ axis.}
    \label{fig:approx_pite_circuit}
\end{figure}

Implementing $\hat{\mathcal U}_{\hat{\mathcal T}}$ requires the exponential operators 
$\exp(\pm i\kappa\hat{\Theta})$, 
where $\kappa\equiv \operatorname{sgn}(\gamma-1/\sqrt2)$ and
$\hat{\Theta}=\arccos\!\left[\left(
\hat{\mathcal T}+\sqrt{1-\hat{\mathcal T}^2}\,\right) /\sqrt{2}
\right]$.
Since a direct implementation of these operators is generally impractical~\cite{kosugi2022}, we expand $\hat{\Theta}$ to first order in the small-$\Delta \tau$ limit, yielding 
\begin{equation}
 \kappa \hat{\Theta}=\theta_0 - \hat{\mathcal{H}} s_1 \Delta \tau+\mathcal{O}\left(\Delta \tau^2\right)\,,
\end{equation}
%$\kappa \hat{\Theta}=\theta_0 - \hat{\mathcal{H}} s_1 \Delta \tau+\mathcal{O}\left(\Delta \tau^2\right)$, 
where $\theta_0 \equiv \kappa \arccos \left[\left(\gamma+\sqrt{1-\gamma^2}\right) / \sqrt{2}\right]$ and $s_1 \equiv \gamma / \sqrt{1-\gamma^2}$.
Consequently, the exponential operator can be implemented using the real-time evolution (RTE) operators
$\hat U_{\mathrm{RTE}}(\Delta t)\equiv e^{-i\hat{\mathcal H}\Delta t}$ with the effective real-time interval $\Delta t = s_1 \Delta \tau$
and single-qubit rotations of the ancilla. 
The resulting approximate PITE circuit, employing the single-qubit gate
$W \equiv \frac{1}{\sqrt{2}}
\left(
\begin{smallmatrix}
1 & -\mathrm{i} \\
1 & \mathrm{i}
\end{smallmatrix}
\right)$\,,
is shown in Fig.~\ref{fig:approx_pite_circuit}, where the forward and backward real-time evolutions are coherently combined.

This circuit represents one imaginary-time step and includes a projective measurement of the ancilla qubit after the step. It is the basis of the package’s shot-based PITE module, where RTE is performed using a Trotter-Suzuki decomposition~\cite{Trotter1959,Suzuki1976}.

Since this shot-based circuit simulation is not the best starting point for simulation and parameter exploration on classical hardware, this package focuses primarily on a state-vector-based formulation of the same approximate PITE update. This formulation avoids explicit ancilla measurements and enables efficient analysis of the evolution and of suitable choices of the parameter $\gamma$ before running shot-based simulations.

\subsection{State-vector formulation}
When simulating the PITE algorithm on classical hardware, it is more convenient to use the state-vector representation of the approximate PITE update~\cite{ejima_probabilistic_2025}, rather than explicitly simulating the ancilla qubit and its projective measurement shown in Fig.~\ref{fig:approx_pite_circuit}. Projecting the ancilla qubit onto the success state $|0\rangle$ yields the following state-vector update. If $|\psi_j \rangle$ is the state vector after the $j$th PITE step, the unnormalised state vector of the system after the $(j+1)$th step is 
\begin{equation}
\left|\psi_{\mathrm{new}}\right\rangle=\frac{1}{2 \sqrt{2}}\left((1-\mathrm{i}) e^{\mathrm{i} \theta_0} \hat{U}_{\mathrm{RTE}}(s_1 \Delta \tau)+(1+\mathrm{i}) e^{-\mathrm{i} \theta_0} \hat{U}_{\mathrm{RTE}}^{\dagger}(s_1 \Delta \tau)\right)\left|\psi_j\right\rangle .
\label{eq:SV_update}
\end{equation}
The normalised state vector after the successful ($j+1$)th step is $|\psi \rangle = |\psi_{\text{new}} \rangle / \sqrt{\mathbb{P}_0^{(j)}}$. Here, $\mathbb{P}_0^{(j)} = \langle \psi_{\text{new}} | \psi_{\text{new}} \rangle$ is the conditional probability of success for the ($j+1$)th step given that the first, second, $\cdots$, and $j$th PITE steps were successful as well. 
This update scheme is implemented in the package's state-vector-based PITE module. 

The state-vector-based PITE simulation does not suffer from statistical uncertainties and corresponds to the infinite-shot limit of the shot-based simulation. Moreover, since it requires neither explicit measurements nor ancillary qubits, the algorithm reduces to a sequence of real-time evolutions. This leads to a substantial improvement in computational efficiency, as demonstrated in later sections using this package.

\section{Package overview}
\label{sec:package_overview}
The purpose of the \SRC{} package is to provide a modular, user-friendly Qiskit-based implementation of the PITE algorithm that integrates into existing workflows. The package implements both the shot‑based approach using a single ancillary qubit and the state-vector‑based formulation. Additionally, it provides an ED interface via QuSpin for direct comparison.

\SRC{} is designed to be easy to use for typical workflows while remaining modular and customizable when needed. Users can define custom initial states, provide alternative backends, or specify custom pass managers for the shot-based algorithm, enabling integration with other quantum simulation pipelines.
Through its interoperability with QuSpin (for ED simulations) and Qiskit, it complements rather than replaces existing tools.

Conceptually, \SRC{} is organized into four main building blocks: Hamiltonian and operator representations, configuration objects, the algorithm classes, and result objects. 
This separation ensures a consistent user interface across simulation methods and allows easy extension.

\subsection{Hamiltonian and operator representation}
All quantum operators in \SRC{} are represented internally as weighted sums of Pauli strings, where each string consists of the Pauli operators \texttt{'X'}, \texttt{'Y'}, and \texttt{'Z'} acting on a fixed number of lattice sites.
The fundamental internal data structure is the \texttt{PauliStringOperator}, which stores the Pauli labels, site indices, and coefficients of each term.
The \texttt{Hamiltonian} class is built on top of \texttt{PauliStringOperator}, enforcing Hermiticity and providing Hamiltonian-specific methods.
This abstraction ensures that both high-level model classes (e.g., Ising and Heisenberg models in one and two dimensions) and user-defined Hamiltonians are expressed in this common internal format used by all three simulation techniques.

Both \texttt{PauliStringOperator} and \texttt{Hamiltonian} instances can be converted to a Qiskit \texttt{SparsePauliOp}, and \texttt{Hamiltonian} instances can additionally be translated into QuSpin Hamiltonians. For more details on converting a  \texttt{PauliStringOperator} or \texttt{Hamiltonian} to other formats, refer to App.~\ref{appsec:api-convert}.

\subsection{Configuration object}
The configuration objects (\texttt{PITEConfig} and \texttt{EDConfig}) are lightweight dataclasses that collect all algorithm parameters (e.g., $\gamma$, $\Delta \tau$, Trotter order, number of steps, shot count, and the choice of initial state) and automatically validate parameter values. 
They make simulations reproducible, provide a common configuration interface across all three algorithms, and help avoid common input mistakes.
Together with the Hamiltonian, the configuration fully specifies the input to each algorithm.

\subsection{Algorithm layer}
Given a Hamiltonian and the corresponding configuration object, the algorithm classes carry out the simulation.
The \texttt{PITEStatevector} and \texttt{PITEShot} classes implement the state-vector-based and shot-based PITE algorithms discussed in Sec.~\ref{sec:PITE}, respectively, while the \texttt{ED} class provides an exact-diagonalisation reference for the PITE simulations.

The \texttt{ED} algorithm class serves as a thin wrapper around the QuSpin package~\cite{weinberg2017}. Using the parameters specified in the \texttt{EDConfig}, QuSpin is employed to construct the appropriate spin-chain basis. The \SRC{} \texttt{Hamiltonian} is then translated into the corresponding QuSpin operator representation, and the ground-state energy and corresponding eigenvector are obtained via QuSpin’s sparse exact-diagonalisation routines.
The resulting ground state serves as a benchmark for the PITE implementation and is used to validate and calibrate PITE parameters.

\subsection{Result objects}
For each algorithm, the simulation output is returned as a dedicated result object, namely, \texttt{PITEStatevectorResult}, \texttt{PITEShotResult}, or \texttt{EDResult}.
These dataclasses store the relevant observables together with the configuration used for the simulation and can optionally include the generated circuit or the computed quantum state.
They provide a consistent interface for analysis and plotting, while allowing additional outputs to be added without changing the user-facing workflow.

\section{Using \SRC: Basic workflow}
\label{sec:basic}

To demonstrate how \SRC{} can be used, this section walks through a standard workflow step by step. 
We assume that the package is installed and that the reader is familiar with basic Python. The installation can be performed easily using the \texttt{uv} package manager, and a detailed guide is provided in App.~\ref{appsec:install}.
We will also assume that all necessary imports are in place. The three primary algorithms are available as top‑level imports, while configurations, results, operators, and Hamiltonian models are imported from dedicated submodules:
\pagebreak
\begin{lstlisting}[style=python]
# ---- Importing everything ----
# Algorithms
from svpite import PITEStatevector, PITEShot, ED

# Configurations
from svpite.configs import PITEConfig, EDConfig

# Operators and Hamiltonians
from svpite.operator import PauliStringOperator, Hamiltonian

# Specific models
from svpite.operator.models import (
    IsingHamiltonian, HeisenbergHamiltonian, 
    XYHamiltonian, XXZHamiltonian
)
\end{lstlisting}
We will consider a concrete physical system and walk through the workflow of obtaining the ground state using state-vector-based PITE, shot-based PITE, and ED, allowing for a comparison of the results. By explaining every step in detail, we aim to familiarize the reader with the usage of the package. 
At the end of the section, we will also demonstrate the usage beyond computing the ground state. The complete code, including the necessary imports and the scripts used to generate the figures, is provided in App.~\ref{appsec:code}.

\subsection{Define the Hamiltonian}

As an example, we consider the $XXZ$ Heisenberg Hamiltonian in a longitudinal magnetic field, defined as
\begin{equation}
\label{eq:H_XXZ_field}
\hat {\cal H} = J \sum_{j=0}^{L-1}
\left( \hat X_j \hat X_{j+1} + \hat Y_j \hat Y_{j+1} + \Delta \hat Z_j \hat Z_{j+1} \right)
+ h_z \sum_{j=0}^{L-1} \hat Z_j\,.
\end{equation}
Here, $\hat X_j$, $\hat Y_j$, and $\hat Z_j$ denote Pauli operators acting on site $j$. The coupling constant $J$ sets the overall interaction scale, $\Delta$ controls the anisotropy, and $h_z$ denotes the strength of the external field in $z$-direction. The system consists of $L$ sites labeled by $j=0,1,\dots,L-1$. We assume periodic boundary conditions (PBC) by identifying site $L$ with site $0$, i.e., $\hat O_L \equiv \hat O_0$ for all local operators $\hat O$.

As a first step, we start by defining the physical model parameters:
\begin{lstlisting}[style=python]
import numpy as np
# ---- Define model parameters ----
L = 8 # number of lattice sites
J = 1.0 / 4.0 # coupling constant
Delta = 1.0 / np.sqrt(2.0) # anisotropy
hz = 1.0 / 5.0 # external field
bc = 'PBC' # periodic boundary conditions
\end{lstlisting}
To define the Hamiltonian $\hat {\cal H}$ from Eq.~\eqref{eq:H_XXZ_field}, we use the predefined $XXZ$ model included in the package as a starting point. 
By passing in the boundary conditions, the two-site terms are automatically constructed in the correct way.
Note that, since operators are internally represented as Pauli strings, all methods for initializing operators use Pauli operators. To work with spin operators, one can  simply use the relation $\hat S^{\alpha}_i = \frac{1}{2}\hat P^{\alpha}_i$ for $\hat P^{\alpha} \in \{ \hat X, \hat Y, \hat Z\}$ and absorbs the corresponding factors of $1/2$ into the model parameters. 
\begin{lstlisting}[style=python]
# ---- Define XXZ Hamiltonian in external z-field ----
# Use standard XXZ model as starting point
H = XXZHamiltonian(L, J, Delta, bc)
\end{lstlisting}
For standard spin models, this is the recommended and easiest way to start. 
Alternatively, especially for non-standard models, one can provide a dictionary of coefficient arrays (\texttt{terms\_dict}) to the \texttt{Hamiltonian} constructor, allowing full control over couplings and lattice structure. Specifically, the dictionary uses strings specifying Pauli operator combinations (e.g., \texttt{"Z"}, \texttt{"XX"}, \texttt{"ZZ"}) as keys and one- or two-dimensional arrays as values, corresponding to single-site or two-site coupling strengths, respectively. The array entries encode the spatial structure of the interaction, i.e., which sites are coupled and with what strength. For two-site terms, the coupling array has the form $J_{ij}$ where each non-zero entry specifies an interaction between site $i$ and site $j$. The $XXZ$ model with PBC can therefore be defined as:
\begin{lstlisting}[style=python]
# Alternatively: Full manual construction
J_xy, J_z = np.zeros((L, L)), np.zeros((L, L))
for i in range(L):
    J_xy[i, (i+1)%L] = J
    J_z[i, (i+1)%L] = J * Delta
terms_dict = {"XX": J_xy, "YY": J_xy, "ZZ": J_z}
H = Hamiltonian(L, terms_dict)
\end{lstlisting}
All Hamiltonian construction methods and further examples are provided in App.~\ref{appsec:api-construct}.
In all cases, the resulting \texttt{Hamiltonian} instance can be passed directly into the PITE algorithm. 
In our example, the external field still needs to be added, which can be done using the \texttt{add\_uniform\_terms} method that adds a translationally invariant term to the Hamiltonian. 
\begin{lstlisting}[style=python]
# Add external z-field terms to the Hamiltonian
H.add_uniform_terms("Z", hz)
\end{lstlisting}
Since we are only adding a single-site term, it is not necessary to specify boundary conditions. In principle, the same term could also be constructed manually, but \texttt{add\_uniform\_terms} provides a more convenient interface for translationally invariant fields.
To verify a correct construction, the Hamiltonian can be displayed using \texttt{print(H)}, yielding:
\begin{align*}
\small
\texttt{0.25\,*}\,\sum_\texttt{i} \texttt{X}_\texttt{i} \texttt{X}_\texttt{i+1} \, \texttt{+\,0.25\,*}\,\sum_\texttt{i} \texttt{Y}_\texttt{i} \texttt{Y}_\texttt{i+1}\,\texttt{+\,0.2\,*}\,\sum_\texttt{i} \texttt{Z}_\texttt{i}\,\texttt{+\,0.177\,*}\,\sum_\texttt{i} \texttt{Z}_\texttt{i} \texttt{Z}_\texttt{i+1}\,\texttt{(Sums\,using\,PBC)}
\end{align*}
As with the whole package, this representation assumes a one-dimensional (1D) spin chain. Two-dimensional (2D) systems are treated by mapping them onto an effective 1D representation. As a result, their printed representation is not easily readable.

\subsection{Algorithm configuration}

Having initialized the Hamiltonian, we can now turn to the configuration of the algorithm. Since the ED implementation is merely a wrapper around standard QuSpin with a Hamiltonian translation layer that is intended primarily for comparison, we focus here on using the two PITE implementations. Some comments on the usage of the ED wrapper are included in Sec.~\ref{sec:comparison_ed}.

Both variants of the PITE algorithm require the same configuration object, \texttt{PITEConfig}, to specify the algorithm parameters. We can therefore initialize a single configuration object and use it for both algorithms in our comparison. The value of \texttt{gamma} used here was determined as described in Sec.~\ref{sec:ground-state}.
\begin{lstlisting}[style=python]
# ---- Define algorithm parameters ----
config = PITEConfig(
    gamma=0.72, # Gamma parameter of the PITE algorithm
    n_steps=80, # Number of imaginary-time evolution steps
    dt=0.2, # Imaginary-time step size
    order=1, # Trotter order for the real-time evolution gate
    reps=1, # Number of Suzuki-Trotter slices per PITE step
    initial_state='singlet', # Initial state of the evolution
    n_shots=10000, # Number of shots for shot-based PITE
)
\end{lstlisting}
Appendix~\ref{appsec:initial_states} details the predefined initial states available in the implementation, which are \texttt{'zero'}, \texttt{'one'}, \texttt{'random'}, \texttt{'plus'}, \texttt{'minus'}, \texttt{'neel'}, \texttt{'neel2d'}, and \texttt{'singlet'}. 
Alternatively, users may provide a Qiskit \texttt{Statevector} (only supported for state-vector-based simulation) or \texttt{QuantumCircuit} to specify a custom initial state with greater overlap with the expected ground state, thereby accelerating convergence.
The parameter \texttt{n\_shots} is optional and is ignored when the configuration is used for a state-vector-based simulation.

\subsection{Running the simulation and accessing the results}
All necessary ingredients for performing the PITE simulations have now been introduced. The interfaces of the shot-based and state-vector-based implementations are nearly identical,
differing only in optional execution parameters.
A simulation is run by initializing the algorithm class with the Hamiltonian and configuration
and then calling the \texttt{run()} method:
\begin{lstlisting}[style=python]
# ---- Initialize and run algorithms ----

# Statevector PITE
sv_algorithm = PITEStatevector(config, H)
sv_result = sv_algorithm.run(verbose=True, return_statevector=True)

# Shot-based PITE
shot_algorithm = PITEShot(config, H)
shot_result = shot_algorithm.run(verbose=True, return_circuit=False)
\end{lstlisting}
By setting \texttt{verbose=True}, the progress of the simulation will be printed during execution. The flag \texttt{return\_statevector} adds the final state-vector to the result object for inspection, comparison or further use, while \texttt{return\_circuit} adds the circuit of the first PITE iteration (including the state preparation) to the result object. This can be useful for analysis or optimisation. 

The shot-based simulation additionally supports the advanced backend options \texttt{backend}, \texttt{use\_primitives}, \texttt{optimization\_level}, \texttt{pass\_manager}, and \texttt{sampler}, which allow fine-grained control over transpilation, execution, and sampling. In the examples below, the default settings are used: \texttt{backend=None} (internally replaced by Qiskit's \texttt{AerSimulator}), \texttt{use\_primitives=True}, and \texttt{optimization\_level=2}. If no custom \texttt{pass\_manager} or \texttt{sampler} is provided, these are generated automatically using the selected backend. These options are not discussed further, as they are part of Qiskit's execution and transpilation framework.

The state-vector-based simulation uses a custom state-vector evolve function and a basic parallelisation strategy by default. This can be changed by passing \texttt{use\_qiskit\_evolve=True} or \texttt{parallel\_evolve=False}, respectively, to the \texttt{run()} function. However, the default settings can provide a speedup of up to $2\times$. For a more detailed discussion of these options and their performance impact, please refer to App.~\ref{appsec:sv_speedup}.

Both calls return result objects that store all relevant observables together with the configuration used for the simulation, and can be printed directly for a short summary. The state-vector result (\texttt{sv\_result}) provides the energy and success probability at each step without sampling noise, whereas the result of the shot-based simulation (\texttt{shot\_result}) only provides the success probability at each step. The energy in the shot-based simulation is measured only after the final step and returned together with its statistical standard deviation.

\begin{lstlisting}[style=python]
# ---- Access and print results ----
print(sv_result)
print(shot_result)

# Statevector PITE results
E_sv = sv_result.energies  # Energy per site vs PITE step
P_sv = sv_result.probabilities  # Success probabilities vs PITE step

# Shot-based PITE results
E_shot = shot_result.energy  # Final energy per site
dE_shot = shot_result.energy_std  # Standard deviation of final energy per site
P_shot = shot_result.probabilities  # Success probabilities vs PITE step
\end{lstlisting}
The state-vector result includes the final state via \texttt{sv\_result.final\_statevector}, which will be used later.
If requested, the first-iteration circuit of the shot-based simulation would be accessible through \texttt{shot\_result.circuit}.

\subsection{Comparison with ED}
\label{sec:comparison_ed}
To establish a reference ground state of our example system, we can use the QuSpin ED integration with the same Hamiltonian $\hat {\cal H}$. The setup closely parallels the PITE workflow, but uses an \texttt{EDConfig} instance instead.
This configuration can specify symmetry sectors through QuSpin basis arguments, as well as optional \texttt{eigsh()} eigensolver parameters:

\begin{lstlisting}[style=python]
# ---- Reference ED calculation ----
# Define ED configuration with symmetry sectors for the ground state computation
ed_basis_kwargs = {"Nup": L//2, "pblock": 1, "kblock": 0}
ed_config = EDConfig (
    tol = 1e-10, # Tolerance for eigensolver
    maxiter = 1e4, # Maximum iterations for eigensolver
    v0 = None, # Initial vector for eigensolver
    spin_basis_1d_kwargs = ed_basis_kwargs
)
\end{lstlisting}
For details on the symmetry sector definitions and eigensolver options, consult the \href{https://quspin.github.io/QuSpin/index.html}{QuSpin documentation}.

Running ED is analogous to running the PITE simulations. The ED call also returns a result object that stores the ground state energy together with the simulation configuration, and optionally the ground-state vector.
\begin{lstlisting}[style=python]
# Initialize and run ED algorithm
ed_algorithm = ED(ed_config, H) 
ed_result = ed_algorithm.run(verbose=True, return_groundstate=False)

# Access ED results
print(ed_result)
E_ed = ed_result.energy # Computed ground state energy
\end{lstlisting}

\begin{figure}[tb]
    \centering
    \includegraphics[width=0.95\linewidth]{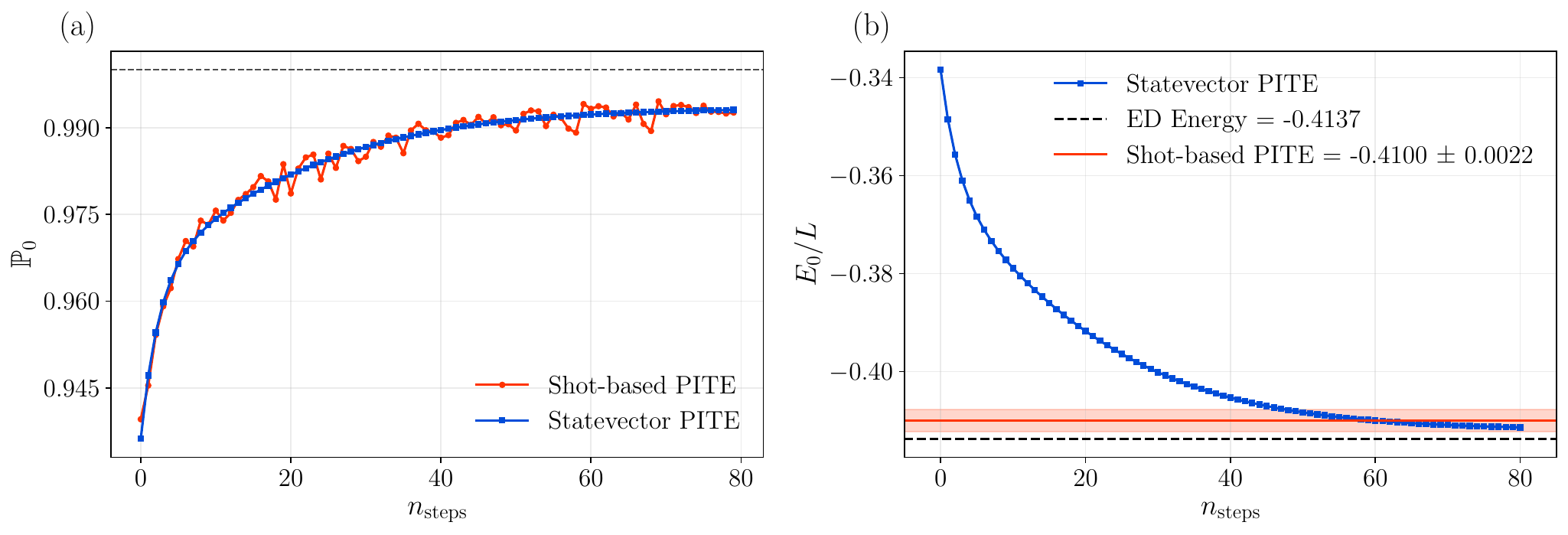}
    \caption{(a) Success probability $\mathbb{P}_0$ as a function of imaginary time steps $n_{\text{steps}}$ for the state-vector-based (blue) and shot-based (red) implementations of the PITE algorithm, obtained using the example code provided in this section. (b) Corresponding energy per site $E_0/L$ as a function of $n_{\text{steps}}$ for the state-vector-based simulation, as well as the final energy estimate from the shot-based PITE simulation including its standard deviation. The exact ground-state energy per site obtained by ED is shown as a black dashed line.}
    \label{fig:basic_workflow_result}
\end{figure}

Figure~\ref{fig:basic_workflow_result} compares the results obtained with the example code using the state-vector-based and shot-based PITE simulations, along with the ED reference.
The computation was performed in approximately 2 minutes on a MacBook Pro with an Apple M1 Max processor. The full code, including the plotting routines, is provided in App.~\ref{appsec:code}.

\subsection{Beyond the basic workflow: Using the \SRC{} Output in further simulations}
\label{subsec:time_evo}
A converged state from the state-vector-based PITE implementation is not limited to ground-state preparation but can serve as the starting point for further simulations. In particular, it can be used within the Qiskit ecosystem or in custom simulation workflows to study, for example, quenches, real-time expectation values, or correlation functions.

The final state returned by \texttt{PITEStatevector} can be directly combined with Qiskit's native state-vector tools and circuit objects. 
For instance, an \SRC{} \texttt{Hamiltonian} can be translated into a Qiskit real-time evolution circuit, allowing one to evolve the prepared state under the same Hamiltonian:

\begin{lstlisting}[style=python]
# Get final statevector from the results
psi0 = sv_result.final_statevector

# Real-time evolution generated by H
U_time = H.to_time_evolution_circuit(time=1, reps=100, order=1)

# Apply the evolution circuit to the prepared state
from svpite import evolve_statevector
psi_t = evolve_statevector(psi0, U_time)
\end{lstlisting}
Here, we used our optimised \texttt{evolve\_statevector()} function instead of the standard Qiskit method. For more details, see App.~\ref{appsec:sv_speedup}.

Expectation values of observables can then be evaluated at different times using standard Qiskit functionality. 
For example, given an operator \texttt{O} in the package representation, one may convert it to a \texttt{SparsePauliOp} and evaluate its expectation value:
\begin{lstlisting}[style=python]
O_qiskit = O.to_sparse_pauli_op()
value_t = psi_t.expectation_value(O_qiskit)
\end{lstlisting}
Repeating this for a sequence of times yields time-dependent observables such as $\langle \hat O(t) \rangle$.

While the Qiskit interface provides a convenient way to continue working with the PITE output, the full wavefunction returned by \texttt{PITEStatevector} can also be accessed directly through \texttt{psi0.data}.
This exposes the underlying NumPy array and enables post-processing outside of Qiskit.
Likewise, operators represented in \SRC{} can not only be converted into Qiskit operators, but also to matrix form:
\begin{lstlisting}[style=python]
# Access full statevector as NumPy array
psi_np = psi0.data

# Convert operator to sparse matrix
O_mat = O.to_matrix(sparse=True)

# Example: expectation value using NumPy
value = np.vdot(psi_np, O_mat @ psi_np)
\end{lstlisting}
We use this functionality in Sec~\ref{sec:DSF} to compute the dynamic structure factor of the Heisenberg model.

\section{Examples and benchmarks}
This section presents a set of application examples that demonstrate how \SRC{} can be used in practical simulations. 
We illustrate ground-state preparation in one- and two-dimensional spin-models, including parameter selection and convergence behaviour. Furthermore, we show how the resulting states enable further studies, including real‑time dynamics and spectral functions. These examples provide a practical reference and benchmark for the PITE implementations. Unless stated otherwise, default parameters were used.

\subsection{Ground-state energy convergence in 1D spin models}
\label{sec:ground-state}

First, we demonstrate how to determine the initial parameter set, $\Delta \tau$ and $\gamma$, for spin models. As an example, we consider the 1D Heisenberg model, which corresponds to Eq.~\eqref{eq:H_XXZ_field} with $\Delta = 1$ and $h_z = 0$.
As before, we assume PBC. Throughout this section, we also set $J = 1/4$ to define the energy unit.

To determine a suitable value $\gamma$, a sweep of state-vector simulations is performed over different values of $\gamma$ for a fixed $\Delta \tau$. 
By repeating this procedure for different choices of $\Delta \tau$ and comparing the results, a suitable set of parameters can be identified.

\begin{figure}[tb]
    \centering
    \includegraphics[width=0.95\linewidth]{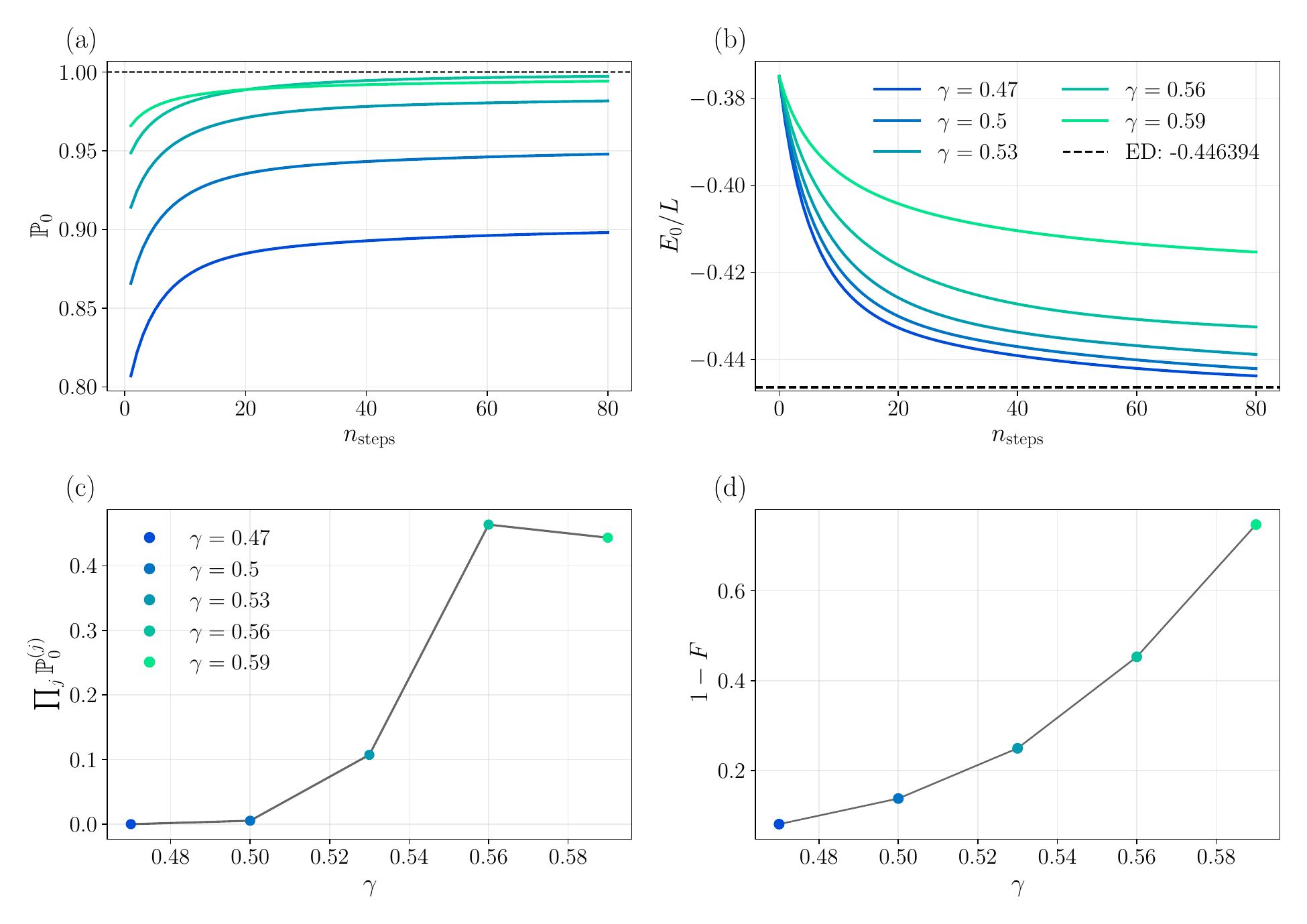}
    \caption{State-vector PITE simulations for different $\gamma$ values for the spin-1/2 Heisenberg chain with $L=16$, using a singlet initial state, and $\Delta \tau = 0.2$. (a) Single-step postselection success probability $\mathbb{P}_0$ and (b) approximate ground-state energy per site $E_0/L$ versus the number of imaginary-time (PITE) steps $n_{\rm steps}$ for each $\gamma$. The exact ground-state energy $E_{\mathrm{ED}}/L$ is shown as a dashed horizontal line. (c) Cumulative success probability $\prod_j \mathbb{P}_0^{(j)}$ and (d) infidelity $1- |\langle\psi_{\mathrm{ED}} | \psi_{\mathrm{svPITE}}\rangle|^2$ versus $\gamma$. The script used to generate this figure is available in \texttt{examples/heisenberg\_gamma\_sweep.py}.}

    \label{fig:heisenberg_gamma_sweep}
\end{figure}

Figure~\ref{fig:heisenberg_gamma_sweep} shows such a sweep for a system with $L=16$ sites and $\Delta \tau = 0.2$. The convergence of the approximate ground-state energy per site $E_0/L$ and success probability $\mathbb{P}_0$ illustrates the trade-off between accuracy and success probability. 
If the exact ground state is available, the infidelity $1 - F = 1 - |\langle\psi_{\mathrm{ED}} | \psi_{\mathrm{svPITE}}\rangle|^2$ provides an additional accuracy measure for the approximate state, complementing the energy-based estimate. 
As shown in panels (b) and (d), the accuracy can be improved by reducing $\gamma$ slightly below $\gamma_{\rm max}$, which is approximately $0.58$ in the present case and yields an almost perfect success probability ($\mathbb{P}_0 \sim 1$), as shown in panel~(a). However, this improved accuracy comes at the cost of a lower success probability, which substantially increases the number of required shots. For instance, in shot-based simulations, the fraction of surviving shots after 80 imaginary-time steps is estimated to be less than 10\% for $\gamma < 0.53$; see panel (c). Thus, in the relevant parameter regime, there is a trade-off between accuracy and success probability, and $\gamma$ should be chosen to achieve an acceptable balance.

\begin{figure}[tb]
    \centering
    \includegraphics[width=0.95\linewidth]{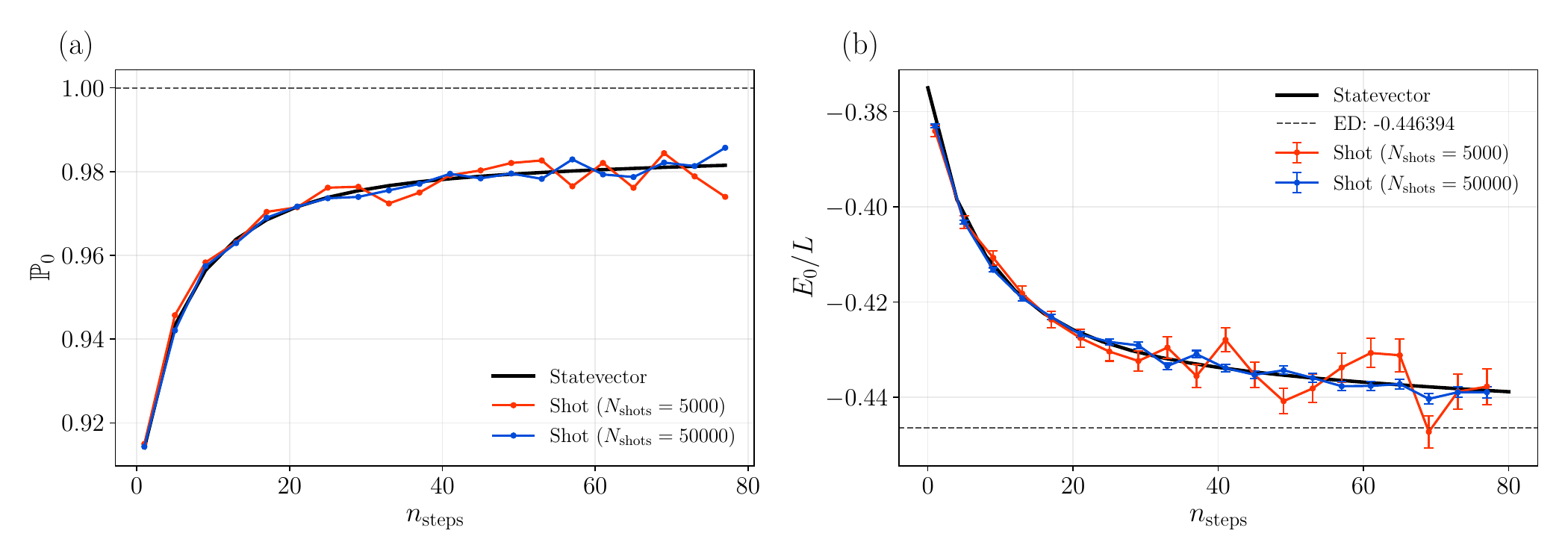}
    \caption{(a) Success probability $\mathbb{P}_0$ and (b) approximate ground-state energy per site $E_0 / L$ as functions of the number of steps, obtained using the shot-based PITE implementation with $\gamma = 0.53$ chosen from Fig.~\ref{fig:heisenberg_gamma_sweep}. The simulations are performed for the spin-1/2 Heisenberg chain with $L=16$, using a singlet initial state and $\Delta \tau = 0.2$. Results obtained with $5{,}000$ shots are shown in red and with $50{,}000$ shots in blue. The solid black line indicates the state-vector PITE reference, while the dashed horizontal line denotes the ED energy. The plot was produced using the scripts in \texttt{examples/heisenberg\_shot\_vs\_sv/}.}
    \label{fig:Heisenberg_Shot_vs_SV}
\end{figure}

For this example, we choose $\gamma = 0.53$ to perform a shot-based simulation, as it offers good accuracy while maintaining a cumulative success probability above $10\%$. 
Figure~\ref{fig:Heisenberg_Shot_vs_SV} compares the results of the shot-based simulations with those of the state-vector simulation. For both the success probability and the energy convergence, we find good agreement between the two approaches. As expected, increasing the number of shots leads to closer agreement with the state-vector result, corresponding to the infinite-shot limit without statistical errors. 
The wall time of the shot-based simulations decreases nearly ideally with the number of cores over the tested range, reflecting the largely independent propagation of the individual shot trajectories.
However, the method remains substantially more expensive than the state-vector approach because many shots are required for statistically reliable estimates, and independent simulations must be performed for different $n_{\mathrm{steps}}$.
A quantitative comparison of the computational cost and parallel scaling is provided in App.~\ref{appsec:shot_performance}.

\pagebreak
\subsection{Ground-state preparation in a 2D lattice model}

The PITE algorithm can be directly applied to higher-dimensional systems, where it may offer potential quantum advantage, particularly in applications to condensed matter physics and materials science.
To illustrate its performance in such settings, we consider the Heisenberg model on a $L\times L$ square lattice with PBC, defined as
\begin{align}   
    \hat {\cal H}_{\rm Heisen2D} = J \sum_{\langle i,j \rangle} 
    \left(
      \hat{X}_i \cdot \hat{X}_j
     +\hat{Y}_i \cdot \hat{Y}_j
     +\hat{Z}_i \cdot \hat{Z}_j
    \right)\,,
    \label{eq:2D-Heisen}
\end{align}
where $\langle i,j \rangle$ denotes nearest-neighbor pairs.

\begin{figure}[tb]
 \centering
 \includegraphics[width=0.95\linewidth]{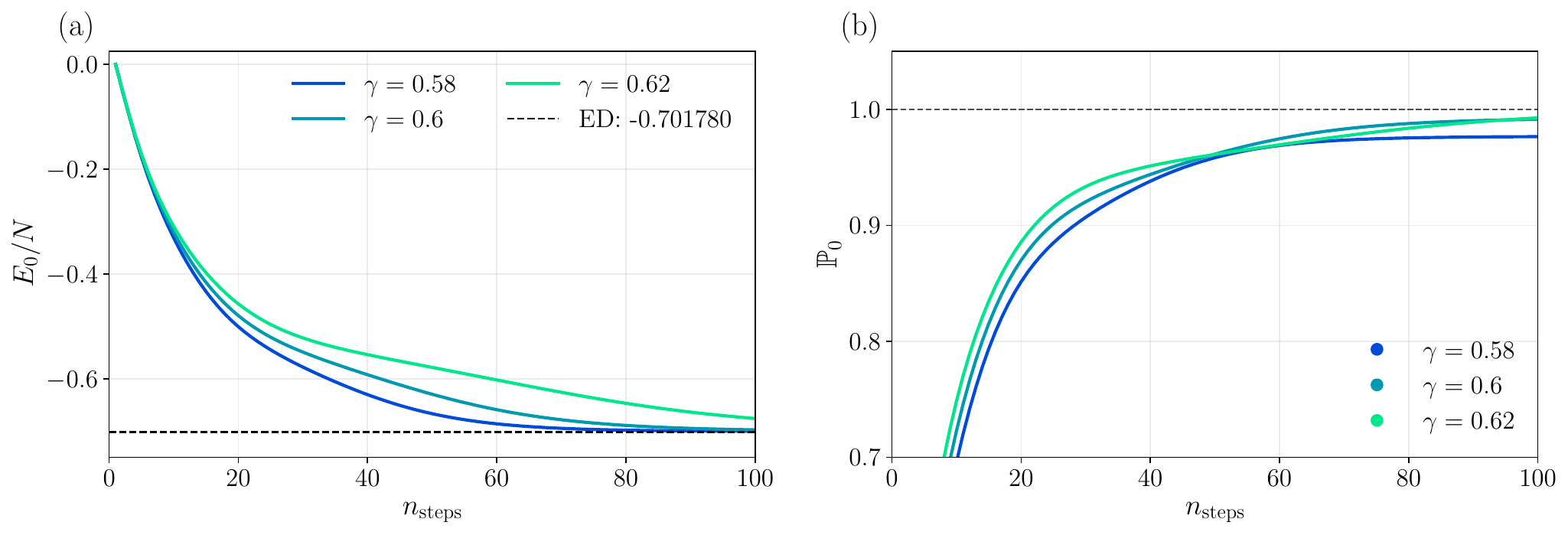}
 \caption{(a) Energy per site $E_0/N$ and (b) success probability $\mathbb{P}_0$ as functions of imaginary-time steps, $n_{\rm step}$, obtained with the state-vector-based PITE algorithm in the 2D Heisenberg model on a $4\times 4$ square lattice (i.e., $N=16$ sites) with PBC. 
 These plots were generated using the script \texttt{examples/heisenberg\_2D\_ed\_sv.py}.}
 \label{fig:2D-Heisen}
\end{figure}

Figure~\ref{fig:2D-Heisen} shows the \SRC{} results for the model~\eqref{eq:2D-Heisen} using $L=4$, i.e., $N=16$ sites. By tuning the initial parameter $\gamma$ to $\gamma=0.6$ for fixed $\Delta\tau = 0.1$, the state evolved from a N\'eel initial state converges to the exact ground state [dashed line in panel (a)] as the number of imaginary-time steps increases. Note that the ground-state energy per site of the $4\times4$ Heisenberg model obtained by ED is $E_0/N=-0.701780$; see, e.g., Ref.~\cite{Sandvik1997}. For these parameter sets, the success probability $\mathbb{P}_0$ remains close to unity, as desired, as shown in Fig.~\ref{fig:2D-Heisen}(b).
For the finite systems considered here, the excitation gap per site is larger in the $4\times4$ cluster than in the $L=16$ chain. This larger finite-size gap is consistent with the faster convergence of PITE in two dimensions.

On fault-tolerant quantum computers with a sufficiently large number of logical qubits, reproducing the high-precision Quantum Monte Carlo (QMC) energies of the 2D square-lattice Heisenberg model by PITE would provide an excellent and nontrivial benchmark. In particular, the recent QMC study of the periodic $L\times L$ system has reported highly accurate finite-size energy data up to $L=96$, offering a valuable reference for assessing the performance of PITE beyond ED system sizes~\cite{Sandvik2026}.

\subsection{Dynamic structure factor of the Heisenberg model}
\label{sec:DSF}

As demonstrated above, the ground state can be systematically determined in both one and two dimensions using \SRC{}. Theoretically, real-time evolution can then be performed on the obtained ground state to compute dynamical correlation functions (e.g., with the help of the Hadamard test). Spectral functions, such as the dynamic structure factor, can be obtained via Fourier transform. However, realizing such simulations on real devices remains challenging due to the required circuit depth. Nevertheless, in this subsection, we provide illustrative examples of spectral functions obtained using \SRC{}.

In 1D spin systems, the dynamic spin structure factor is defined as 
\begin{align}
 S(q,\omega)=\int_{-\infty}^{\infty}dt \sum_r 
  e^{\mathrm{i}(qr-\omega t)}
  \langle{\hat{S}_{j+r}^{z}(t)\hat{S}_{j}^{z}(0)}\rangle\,.
 \label{eq:Sqw}
\end{align}
In the following, we describe the computational procedure for computing $S(q,\omega)$ using \SRC{}, step by step.
First, we prepare the ground state $|\psi_0\rangle$ by PITE or ED algorithms as explained in Sec.~\ref{sec:ground-state}. 
In order to compute the time-dependent correlation functions in Eq.~\eqref{eq:Sqw}, we then act with a local spin operator $\hat{S}_j^z$ on the ground state and prepare the initial state $|\phi(0)\rangle=\hat{S}_j^z|\psi_0\rangle$. This state is evolved in real time as
$|\phi(t)\rangle=e^{-\mathrm{i}\hat{\cal H} t}\hat{S}_j^z|\psi_0\rangle$, where $t=n\Delta t$ and $\Delta t$ is a small time step. (See Sec.~\ref{subsec:time_evo} for the time-evolution procedure using \texttt{evolve\_statevector()} in this package.) The dynamical spin correlation function is then evaluated as
\begin{align}
    C^{zz}(r, t) = e^{\mathrm{i}E_0 t} 
       \langle{\psi_0|\hat{S}_{j+r}^{z} e^{-\mathrm{i}\hat{\cal H} t} \hat{S}_{j}^{z}}|\psi_0\rangle\,.
\end{align}
Finally, the momentum- and frequency-resolved dynamical structure factor, $S(q,\omega)$, is obtained by performing the space-time Fourier transform of $C^{zz}(r,t)$.

\begin{figure}[tb]
    \centering
    \includegraphics[width=1.0\linewidth]{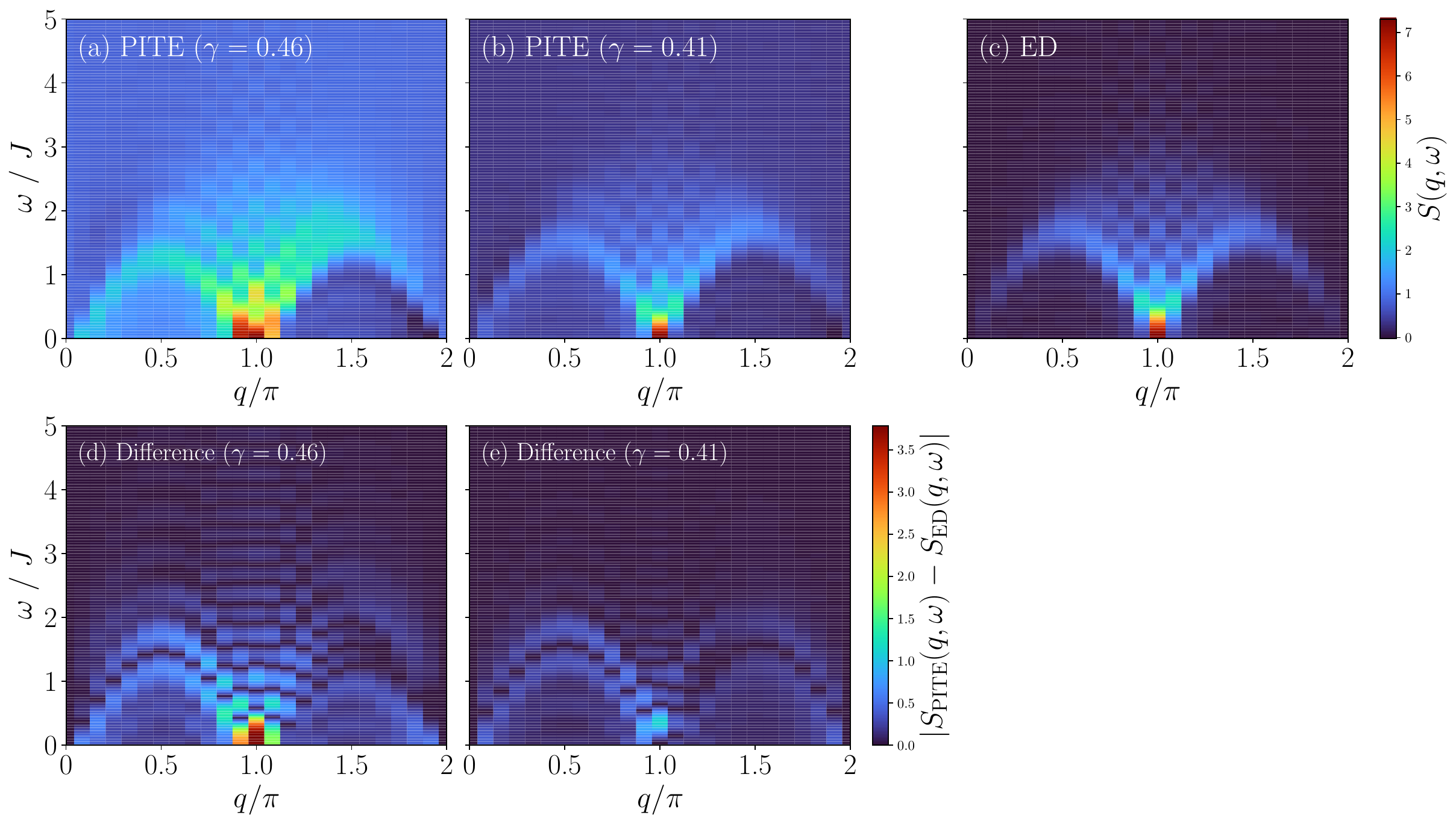}
    \caption{Dynamic structure factor $S(q,\omega)$ with $L=24$ and PBC in the Heisenberg model, obtained using \SRC{}. Panels (a) and (b) show results obtained using the state-vector-based PITE algorithm with $\gamma=0.46$ and $0.41$, respectively, while panel (c) shows results obtained using the ED algorithm. Panels (d) and (e) show $|S_{\mathrm{PITE}}(q,\omega)-S_{\mathrm{ED}}(q,\omega)|$ for $\gamma=0.46$ and $0.41$, respectively, with a common color scale.
    }
    \label{fig:dsf}
\end{figure}

Figure~\ref{fig:dsf} shows the obtained results for $S(q,\omega)$ in the Heisenberg chain, corresponding to the $XXZ$ Hamiltonian in Eq.~\eqref{eq:H_XXZ_field} with $\Delta = 1$ and $h_z =0$, for system size $L=24$ and PBC. Starting from the ground state obtained with the optimal $\gamma=0.46$ for fixed $\Delta\tau=0.2$ and evolving the system in real time with the time step of $\Delta t=0.025$, $S(q,\omega)$ becomes asymmetric due to the inaccuracy of the ground state as shown in Fig.~\ref{fig:dsf}(a), although it should be symmetric about $q=\pi$ as demonstrated by the ED results in Fig.~\ref{fig:dsf}(c). 
Using a more accurate ground state obtained with $\gamma=0.41$ substantially reduces this asymmetry, as shown in Fig.~\ref{fig:dsf}(b).
To quantify this improvement, Figs.~\ref{fig:dsf}(d) and (e) show the absolute differences between the svPITE and ED results. The deviation is visibly reduced for $\gamma=0.41$ compared with $\gamma=0.46$, confirming the improved accuracy of the dynamical structure factor. However, the success probability after 200 imaginary-time steps decreases from 99.9\% for $\gamma=0.46$ to 96.3\% for $\gamma=0.41$, indicating that a larger number of shots would be required in shot-based simulations.

We would like to note that the code for reproducing Fig.~\ref{fig:dsf} is available as \texttt{examples\allowbreak /heisenberg\_RTE\_ed.py} and \texttt{examples/heisenberg\_RTE\_sv.py}, for the ED and state-vector PITE algorithms, respectively.

\section{Future perspectives}

Several directions for future development could further broaden the applicability and performance of the \SRC{} package.

A first important extension concerns the treatment of more general interaction structures. At present, the software is well suited to non-mixed interaction terms up to the two-body level, but many physically relevant models involve mixed interaction terms or higher-order many-body contributions. Supporting mixed interaction terms in a more flexible and automated manner would significantly enlarge the class of systems that can be studied within a unified workflow. In particular, a generalized term-handling interface could simplify the implementation of composite Hamiltonians and make the framework more useful for realistic many-body applications.

A second major direction concerns computational efficiency and backend flexibility. In particular, a more detailed investigation is needed into the extent to which efficient parallelisation can be achieved for state-vector-based simulations at larger qubit numbers, where the computational overhead grows rapidly. Clarifying this point will be important for assessing the practical scalability of the current implementation. Although the current implementation relies on Qiskit, it would also be of interest to examine alternative platforms such as Qrisp~\cite{qrisp}, Cirq/qsim~\cite{cirq-qsim}, Qulacs~\cite{Qulacs}, PennyLane Lightning~\cite{PennyLane}, or \href{https://github.com/NVIDIA/cuda-quantum}{CUDA-Q}, and to assess how these choices influence the efficiency and practicality of parallelised state-vector-based simulations at larger qubit numbers.

From a physics perspective, an especially relevant next step is the extension of the present framework to spin systems at finite temperature. While the current implementation focuses on ground-state-oriented or zero-temperature settings, many important questions in condensed-matter and statistical physics require access to thermal observables and finite-temperature states. Incorporating suitable purification schemes, thermal-state preparation strategies, or imaginary-time techniques adapted to mixed states would open the way to a wider range of applications. 

Another natural target is the study of Fermi–Hubbard models. These models are among the central benchmarks in quantum simulation and are directly relevant for strongly correlated fermionic systems. However, extending PITE towards fermionic lattice Hamiltonians would require robust support for fermion-to-qubit mappings, efficient treatment of nonlocal operator strings, and careful optimisation of measurement costs. Achieving this would considerably strengthen the scientific scope of the software and connect it to a broad body of ongoing work in quantum many-body physics.

Overall, future development should proceed along two complementary lines: expanding the range of physical models to include mixed interactions, finite-temperature spin systems, and fermionic models such as the Fermi–Hubbard Hamiltonian, while also improving the software architecture through enhanced parallelisation on both traditional HPC systems and GPU-based platforms. Together, these developments would make \SRC{} a more versatile and sustainable platform for quantum simulation research.

% In the list of references, cited papers \cite{1931_Bethe_ZP_71} should include authors, title, journal reference (journal name, volume number (in bold), start page) and most importantly a DOI link. For a preprint \cite{arXiv:1108.2700}, please include authors, title (please ensure proper capitalization) and arXiv link. If you use BiBTeX with our style file, the right format will be automatically implemented.

% All equations and references should be hyperlinked to ensure ease of navigation. This also holds for [sub]sections: readers should be able to easily jump to Section \ref{sec:another}.

% \section{Conclusion}
% Conclusion here.

\section*{Acknowledgements}
SE acknowledges the Q-Neko project, which has received funding from the European Union’s Horizon Europe research and innovation programme under Grant Agreement No. 101241875. This work was also performed for Council for Science, Technology and Innovation (CSTI), Cross-ministerial Strategic Innovation Promotion Program (SIP), “Promoting the application of advanced quantum technology platforms to social issues”(Funding agency: QST).

This project was made possible by the DLR Quantum Computing Initiative and the Federal Ministry for Economic Aﬀairs and Climate Action; qci.dlr.de/projects/ALQU.

The authors gratefully acknowledge the scientific support and HPC resources provided by the German Aerospace Center (DLR). The HPC system CARO is partially funded by "Ministry of Science and Culture of Lower Saxony" and "Federal Ministry of Research, Technology and Space".

% \paragraph{Author contributions}
% This is optional. If desired, contributions should be succinctly described in a single short paragraph, using author initials.

% TODO: include funding information
% \paragraph{Funding information}
% Authors are required to provide funding information, including relevant agencies and grant numbers with linked author's initials. Correctly-provided data will be linked to funders listed in the \href{https://www.crossref.org/services/funder-registry/}{\sf Fundref registry}.

\newpage

\begin{appendix}
\numberwithin{equation}{section}

\section{Installation guide}
\label{appsec:install}

\SRC{} and its dependencies can be installed using the package and project manager \texttt{uv}.
The package depends on \texttt{QuSpin} and \texttt{Qiskit} and is tested with Python \texttt{3.11} and \texttt{3.12}. All required dependencies are specified in the project configuration \texttt{pyproject.toml} and are installed automatically (including a compatible Python version) when following the minimal installation procedure below. 
The use of \texttt{uv} ensures a simplified setup of reproducible environments and provides unified dependency and Python version management.

\subsection{Installing \texttt{uv}}
\texttt{uv} can be installed as a standalone binary without requiring an existing Rust or Python installation.  
For more detailed information, refer to the official \href{https://docs.astral.sh/uv/}{\texttt{uv} documentation}.

\subsubsection{MacOS / Linux}
\begin{lstlisting}[style=shell]
 $ curl -LsSf https://astral.sh/uv/install.sh | sh
 $ uv --version
\end{lstlisting}
Alternatively, \texttt{uv} can be installed from \href{https://pypi.org/project/uv/}{PyPI} using \texttt{pip} or \texttt{pipx}, or via \href{https://formulae.brew.sh/formula/uv}{Homebrew} on macOS. If installed using the standalone script, \texttt{uv} can be updated by running \texttt{uv self update}; otherwise, updates are handled by \texttt{pip} or \texttt{brew}.

\subsubsection{Windows}
\begin{lstlisting}[style=shell]
 > powershell -ExecutionPolicy ByPass -c "irm https://astral.sh/uv/install.ps1 | iex"
 > uv --version
\end{lstlisting}

\subsection{Installing \SRC{}}
The installation procedure is identical on Windows, macOS, and Linux.  
Clone the repository and run \texttt{uv sync} to install all dependencies:
\begin{lstlisting}[style=shell]
 $ git clone https://gitlab.com/dlr-sc-qc/many-body/svpite
 $ cd svpite
 $ uv sync
\end{lstlisting}
This creates a dedicated virtual environment \texttt{.venv} inside the package directory and installs all dependencies specified in \texttt{pyproject.toml}. Existing Python installations are not affected.

To install additional dependencies required for some example scripts and notebooks:
\begin{lstlisting}[style=shell]
$ uv sync --group examples
\end{lstlisting}
More detailed usage guidance for \texttt{uv} (e.g., adding packages, updating environments) is available in the project \texttt{README} and in the \texttt{uv} documentation.

Any Python script can now be executed using \texttt{uv run}:
\begin{lstlisting}[style=shell]
$ uv run examples/ising_4_sv.py
\end{lstlisting}

Alternatively, the virtual environment can be activated manually:
\begin{lstlisting}[style=shell]
# On macOS and Linux
$ source .venv/bin/activate
\end{lstlisting}
\begin{lstlisting}[style=shell]
# On Windows
> .\.venv\Scripts\activate
\end{lstlisting}
When the environment is active, Python commands and package installations affect only this environment.

\section{Predefined initial states}
\label{appsec:initial_states}
The predefined initial states supported by \texttt{PITEConfig} are:

\begin{itemize}
    \item \texttt{'zero'}: the all-zero computational basis state
    \[
    |0\rangle^{\otimes L}.
    \]
    
    \item \texttt{'one'}: the all-one computational basis state
    \[
    |1\rangle^{\otimes L}.
    \]
    
    \item \texttt{'plus'}: the product state with every qubit in the $|+\rangle$ state,
    \[
    |+\rangle^{\otimes L},
    \qquad |+\rangle=\frac{|0\rangle+|1\rangle}{\sqrt{2}}.
    \]
    
    \item \texttt{'minus'}: the product state with every qubit in the $|-\rangle$ state,
    \[
    |-\rangle^{\otimes L},
    \qquad |-\rangle=\frac{|0\rangle-|1\rangle}{\sqrt{2}}.
    \]
    
    \item \texttt{'singlet'}: a product of two-qubit singlet states prepared on neighbouring pairs of sites,
    \[
    \bigotimes_{j=1}^{L/2}
    \frac{|01\rangle_{2j-1,2j}-|10\rangle_{2j-1,2j}}{\sqrt{2}}.
    \]
    This option requires an even number of sites.
    
    \item \texttt{'neel'}: the one-dimensional Néel state with alternating spin configuration,
    \[
    |1010\ldots\rangle,
    \]
    where every second qubit is flipped relative to the all-zero state.
    
    \item \texttt{'neel2d'}: the two-dimensional Néel state on a square lattice, where neighbouring sites have opposite spin orientations. This option requires \texttt{n\_sites} to be a perfect square. If the lattice has an odd number of sites along each side, the resulting state coincides with the one-dimensional Néel state.
    
    \item \texttt{'random'}: a random product state obtained by preparing each qubit independently in a state of the form
    \[
    H R_z(\phi) H |0\rangle,
    \]
    where each angle \(\phi\) is sampled independently and uniformly from \([0,2\pi)\).
\end{itemize}

\section{Hamiltonian and operator method reference}
\label{appsec:api}

This appendix provides a detailed explanation of the complete signatures and behaviour of the user-facing methods of the \texttt{PauliStringOperator} and \texttt{Hamiltonian} classes introduced in Sec.~\ref{sec:package_overview}, together with the four predefined model classes that are subclasses of \texttt{Hamiltonian}. Unless stated otherwise, a method is described for \texttt{PauliStringOperator} and is inherited unchanged by \texttt{Hamiltonian} and all model subclasses; methods marked ``\texttt{Hamiltonian} only'' are not available on a plain \texttt{PauliStringOperator}.

\subsection{Constructing and modifying operators}
\label{appsec:api-construct}

Both \texttt{PauliStringOperator} and \texttt{Hamiltonian} share the same constructor signature. However, for a \texttt{Hamiltonian}, Hermiticity is checked after every addition of terms.

We illustrate different ways of constructing the same Hamiltonian using the Ising model as an example:
\begin{equation*}
\hat {\cal H} = J \sum_{\langle i,j \rangle} \hat Z_i \hat Z_j + h \sum_i \hat X_i\,.
\end{equation*}
All construction examples shown here for \texttt{Hamiltonian} also work for \texttt{PauliStringOperator}.

We assume that all necessary imports are already in place. The boundary condition parameter \texttt{bc} is either \texttt{'PBC'} or \texttt{'OBC'}, corresponding to periodic and open boundary conditions, respectively. The parameters $J$ and $h$ are scalar coupling coefficients of type \texttt{float}.

\subsubsection{Using a predefined model}
If the desired model is among the available predefined models, using it is the easiest approach:
\begin{lstlisting}[style=python]
H = IsingHamiltonian(n_sites=4, J=J, h=h, bc=bc)
\end{lstlisting}
All available predefined models are outlined in Sec.~\ref{appsec:api-models}.

\subsubsection{Using a \texttt{terms\_dict}}
One approach to defining an operator is to define a \texttt{terms\_dict} dictionary. The following definitions all produce the same operator:

\begin{lstlisting}[style=python]
# Translationally invariant Hamiltonian from coefficients
terms_dict = {"ZZ": (J, bc), "X": h}

# Coefficient arrays
J_zz = np.zeros((4,4))
for i in range(4):
    J_zz[i, (i+1)%4] = J
h_x = np.array([h for _ in range(4)])
terms_dict = {"ZZ": J_zz, "X": h_x}

# List of [coeff, sites] pairs
zz_terms = [[J, (i, (i+1)%4)] for i in range(4)]
x_terms = [[h, (i,)] for i in range(4)]
terms_dict = {"ZZ": zz_terms, "X": x_terms}
\end{lstlisting}

The \texttt{terms\_dict} can now be passed directly to the constructor or added to an already existing empty operator:
\begin{lstlisting}[style=python]
# Construct using terms_dict
H = Hamiltonian(n_sites=4, terms_dict=terms_dict)
\end{lstlisting}
\begin{lstlisting}[style=python]
# Add terms_dict after construction
H = Hamiltonian(n_sites=4)
H.add_terms_dict(terms_dict)
\end{lstlisting}

\subsubsection{Term-adding methods}
Alternatively, terms can also be added directly to the \texttt{Hamiltonian} without first defining a \texttt{terms\_dict}. In the following examples, we assume that \texttt{H} is an existing empty \texttt{Hamiltonian} with four sites.
\begin{lstlisting}[style=python]
# Construction with add_term
for i in range(4):
    H.add_term("ZZ", J, (i, (i+1)%4))
    H.add_term("X", h, (i,))
    
# Construction with add_uniform_terms
H.add_uniform_terms("ZZ", J, bc)
H.add_uniform_terms("X", h)

# Manual construction from coefficient arrays
J_zz = np.zeros((4,4))
for i in range(4):
    J_zz[i, (i+1)%4] = J
h_x = np.array([h for _ in range(4)])
H.add_array_terms("ZZ", J_zz)
H.add_array_terms("X", h_x)

# Manual construction using a list of [coeff, sites] pairs
zz_terms = [[J, (i, (i+1)%4)] for i in range(4)]
x_terms = [[h, (i,)] for i in range(4)]
H.add_coeff_list_terms("ZZ", zz_terms)
H.add_coeff_list_terms("X", x_terms)
\end{lstlisting}

\subsubsection{Shaped 2D operators}
As outlined in the main text, all operators are internally represented as Pauli strings. However, to simplify working with 2D models, operators and Hamiltonians can also be defined as
\begin{lstlisting}[style=python]
H = Hamiltonian(shape=(n_rows, n_cols))
\end{lstlisting}
where \texttt{n\_rows} is the number of rows and \texttt{n\_cols} is the number of columns of the 2D lattice.

For such a shaped operator, all methods remain the same, but site indices can be provided as coordinate tuples instead of flattened integer indices. Flattened (row-major) integer indices are still supported, but we recommend using coordinate tuples instead. The \texttt{add\_uniform\_terms} method applies one-site labels to every lattice site and two-site labels to all horizontal and vertical nearest-neighbour bonds of the rectangular 2D lattice. Coefficient arrays are unaffected and continue to use flat integer indices. Several examples illustrating the construction of shaped Hamiltonians are given in App.~\ref{appsubsec:2D_Hamiltonian_examples}.

\subsection{Predefined models}
\label{appsec:api-models}
Four commonly used spin-lattice models are provided as ready-made subclasses of \texttt{Hamiltonian}: the transverse-field Ising model, the Heisenberg model, the XY model, and the XXZ model. All models support both 1D chains (by specifying \texttt{n\_sites}) and 2D rectangular lattices (by giving the \texttt{shape} tuple).  Two-site interactions automatically connect nearest neighbours, with boundary conditions set by the string argument \texttt{bc} (\texttt{'PBC'} or \texttt{'OBC'}).  All methods inherited from \texttt{Hamiltonian}---term manipulation, conversion to other representations, and circuit generation---work without any further configuration.

Each predefined model requires one or two additional coupling constants, summarised below together with its Hamiltonian:

\begin{description}
\item[IsingHamiltonian] \hfill \\
\(\displaystyle
\hat{\mathcal H} = J \sum_{\langle i,j \rangle} \hat Z_i \hat Z_j + h \sum_i \hat X_i .
\)
The constructor accepts \texttt{J} and \texttt{h}.

\item[HeisenbergHamiltonian] \hfill \\
\(\displaystyle
\hat{\mathcal H} = J \sum_{\langle i,j \rangle} \bigl(\hat X_i \hat X_j + \hat Y_i \hat Y_j + \hat Z_i \hat Z_j\bigr) .
\)
Only the coupling \texttt{J} is required.

\item[XYHamiltonian] \hfill \\
\(\displaystyle
\hat{\mathcal H} = J \sum_{\langle i,j \rangle} \bigl(\hat X_i \hat X_j + \hat Y_i \hat Y_j\bigr) .
\)
Only the coupling \texttt{J} is required.

\item[XXZHamiltonian] \hfill \\
\(\displaystyle
\hat{\mathcal H} = J \sum_{\langle i,j \rangle} \bigl(\hat X_i \hat X_j + \hat Y_i \hat Y_j + \Delta\, \hat Z_i \hat Z_j\bigr) .
\)
The constructor expects \texttt{J} and \texttt{Delta}.
\end{description}

Constructing a model follows the same pattern as a generic \texttt{Hamiltonian}:
\begin{lstlisting}[style=python]
H = IsingHamiltonian(n_sites=6, J=1.0, h=0.5, bc='PBC')
H2d = HeisenbergHamiltonian(shape=(3, 4), J=-1.0, bc='OBC')
\end{lstlisting}
Once built, the operator can be inspected, modified, or passed directly to the time-evolution and conversion methods described in Secs.~\ref{appsec:api-convert} and~\ref{appsec:api-circuits}.

\subsection{Converting to other representations and packages}
\label{appsec:api-convert}

An operator can be freely moved between \SRC{}'s own representations and the external packages it interfaces with. This includes switching between \texttt{PauliStringOperator} and \texttt{Hamiltonian}, exporting to Qiskit, and exporting to QuSpin.

\subsubsection{Switching between \texttt{PauliStringOperator} and \texttt{Hamiltonian}}
Since a \texttt{Hamiltonian} checks Hermiticity after every added term, it can be convenient to first assemble an operator as a plain \texttt{PauliStringOperator}, for example when combining several individually non-Hermitian pieces, and only promote it once construction is complete:
\begin{lstlisting}[style=python]
op = PauliStringOperator(n_sites=4)
op.add_term("XY", 1j * g, (0, 1))
op.add_term("YX", 1j * g, (0, 1))
...
H = op.to_hamiltonian()   # checks Hermiticity once, then returns a Hamiltonian
\end{lstlisting}
The reverse conversion (\texttt{Hamiltonian} only) drops the automatic Hermiticity enforcement for any subsequent modifications:
\begin{lstlisting}[style=python]
op = H.to_pauli_string_operator()
\end{lstlisting}

\subsubsection{Qiskit and NumPy/SciPy representations}
Any \texttt{PauliStringOperator} or \texttt{Hamiltonian} can be exported as a Qiskit \texttt{SparsePauliOp}, or directly as a matrix:
\begin{lstlisting}[style=python]
O_qiskit = O.to_sparse_pauli_op()

O_dense  = O.to_matrix(sparse=False)   # numpy.ndarray
O_sparse = O.to_matrix(sparse=True)    # scipy sparse matrix
\end{lstlisting}
These are the same conversions used in Sec.~\ref{subsec:time_evo} to evaluate observables on a state produced by the state-vector PITE simulation.

\subsubsection{QuSpin representation (\texttt{Hamiltonian} only)}
For exact diagonalisation or other QuSpin-based post-processing, a \texttt{Hamiltonian} can be exported to a QuSpin \texttt{hamiltonian} object:
\begin{lstlisting}[style=python]
H_quspin = H.to_quspin_hamiltonian(**spin_basis_1d_kwargs)
\end{lstlisting}
Keyword arguments are forwarded to QuSpin's \texttt{spin\_basis\_1d} (see the \href{https://quspin.github.io/QuSpin/index.html}{QuSpin documentation}), so symmetry sectors can be specified the same way as for the \texttt{EDConfig} used by the \texttt{ED} algorithm (Sec.~\ref{sec:comparison_ed}), which relies on this conversion internally. QuSpin itself only ever sees a 1D chain of \texttt{n\_sites} sites; since all couplings are already stored via flattened site indices (Sec.~\ref{appsec:api-construct}), this conversion works identically for 1D chains and for shaped 2D Hamiltonians.

\subsection{Generating time-evolution circuits (\texttt{Hamiltonian} only)}
\label{appsec:api-circuits}

 Real-time evolution Qiskit \texttt{QuantumCircuit} implementing (approximate) unitaries $e^{-i\hat{\cal H}\Delta t}$ are built from a Qiskit \texttt{PauliEvolutionGate} together with a Suzuki--Trotter synthesis, and can be generated directly from a \texttt{Hamiltonian} instance.

\subsubsection{Time-evolution circuit}
\begin{lstlisting}[style=python]
U_time = H.to_time_evolution_circuit(time, reps=100, order=1)
\end{lstlisting}
returns a Qiskit \texttt{QuantumCircuit} on \texttt{n\_sites} qubits implementing $e^{-i\hat{\cal H}t}$, where \texttt{time} is the total evolution time $t$, \texttt{reps} is the number of Trotter steps used to subdivide \texttt{time} (larger \texttt{reps} gives a finer approximation), and \texttt{order} selects the order of the Suzuki--Trotter decomposition, e.g.\ \texttt{order=1} for the first-order product formula or \texttt{order=2} for the symmetric second-order formula. This is the circuit used to continue a PITE simulation with real-time evolution in Sec.~\ref{subsec:time_evo}.

\subsubsection{Controlled time-evolution circuit}
\begin{lstlisting}[style=python]
cU_time = H.to_controlled_time_evolution_circuit(time, reps=100, order=1)
\end{lstlisting}
builds the same Trotterized evolution circuit as above for the total evolution time \texttt{time}, but returns it as a single ancilla-controlled gate acting on \texttt{n\_sites}\,+\,1 qubits. This is the building block used to construct the ancilla-controlled forward and backward evolution operators of the shot-based PITE circuit shown in Fig.~\ref{fig:approx_pite_circuit}.

\subsection{Hermiticity checks, comparison, and utility methods}
\label{appsec:api-utility}

Beyond term construction, a handful of methods are available for checking, comparing, and inspecting operators. All are shared between \texttt{PauliStringOperator} and \texttt{Hamiltonian}.

\subsubsection{Hermiticity}
As noted in Sec.~\ref{appsec:api-construct}, term-adding methods automatically check Hermiticity for \texttt{Hamiltonian} instances after every call. The same check can also be run explicitly, e.g.\ on a \texttt{PauliStringOperator} before promoting it, or on demand for a \texttt{Hamiltonian}:
\begin{lstlisting}[style=python]
op.is_hermitian()        # bool, does not raise
op.check_hermiticity()   # raises ValueError and prints offending terms if not Hermitian
\end{lstlisting}
The Hermitian conjugate of an operator can be obtained as a new object, or applied in place:
\begin{lstlisting}[style=python]
op_dag = op.hermitian_conjugate()
op.hermitian_conjugate(inplace=True)
\end{lstlisting}

\subsubsection{Comparing and combining operators}
Two operators compare equal if they act on the same number of sites and contain the same terms, independent of the order in which the terms were added; a \texttt{Hamiltonian} and a \texttt{PauliStringOperator} with identical terms are never equal to one another:
\begin{lstlisting}[style=python]
op1 == op2
\end{lstlisting}
Operators can also be added directly with \texttt{+}, provided they act on the same number of sites (and, if shaped, on the same \texttt{shape}). Adding two \texttt{Hamiltonian}s returns a \texttt{Hamiltonian}, adding two \texttt{PauliStringOperator}s returns a \texttt{PauliStringOperator}, and mixing the two types raises a \texttt{TypeError}:
\begin{lstlisting}[style=python]
H_total = H1 + H2
\end{lstlisting}

\subsubsection{Inspecting and modifying operators}
All terms of a given Pauli label can be dropped at once:
\begin{lstlisting}[style=python]
H.remove_pauli_label("ZZ")
\end{lstlisting}
For a quick, human-readable overview of an operator, \texttt{print(H)} gives a short summary, while \texttt{H.show()} additionally prints one line per term, marking the sites it acts on together with its coefficient which is convenient for sanity-checking a newly constructed Hamiltonian.

\section{Parallelisation and optimisation of state-vector simulations}
\label{appsec:sv_speedup}

To optimise the state-vector-based PITE simulation and thereby accelerate parameter exploration, we implemented two main strategies: (i) parallel simulation of the real-time evolution $\hat{U}_{\mathrm{RTE}}$ and its Hermitian adjoint $\hat{U}_{\mathrm{RTE}}^{\dagger}$, and (ii) a custom state-vector evolution routine tailored specifically to the operators used in the PITE algorithm. 
By combining these two optimisations, we achieve an overall speedup of approximately $2\times$ for systems with $L \geq 22$ sites.

All benchmark results presented in this section were obtained using the scripts available in \texttt{examples/sv\_evolve\_benchmarks/}.

\subsection{Parallelisation of real-time evolution}

Computing the updated state-vector according to Eq.~\eqref{eq:SV_update} involves a linear combination of forward- and backward-time evolution operators acting on the current state. The application of these two operators can therefore be evaluated in parallel. This behaviour can be controlled via the \texttt{parallel\_evolve} parameter of the state-vector \texttt{run()} function. It is enabled by default, as it provides a speedup for systems larger than $L \gtrsim 12$. 

\begin{figure}[H]
 \centering
 \includegraphics[width=0.95\linewidth]{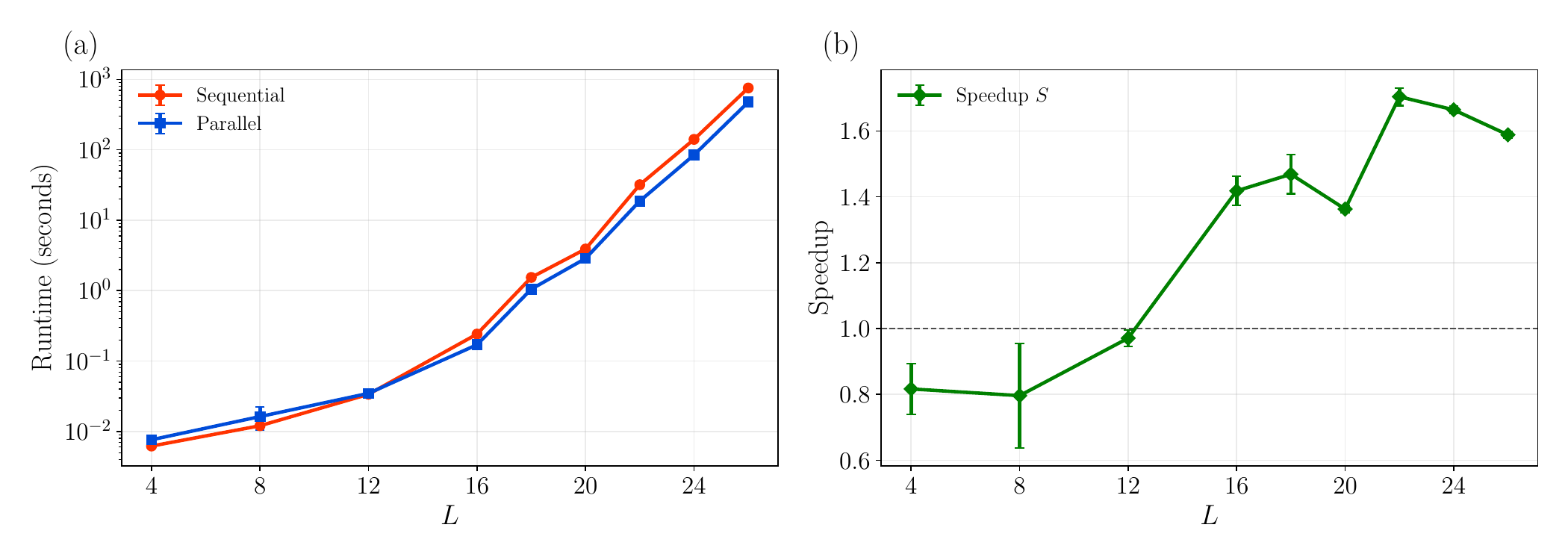}
 \caption{(a) Runtime of state-vector PITE simulations using sequential and parallel evolution as a function of system size $L$ for a fixed number of steps, $n_{\rm steps} = 10$. (b) Corresponding speedup, $S = t_{\rm seq} / t_{\rm par}$. The data correspond to a Heisenberg chain. Error bars show the standard deviation over 10 simulations per data point.}
 \label{fig:parallel_speedup2}
\end{figure}

\begin{figure}[H]
 \centering
 \includegraphics[width=0.95\linewidth]{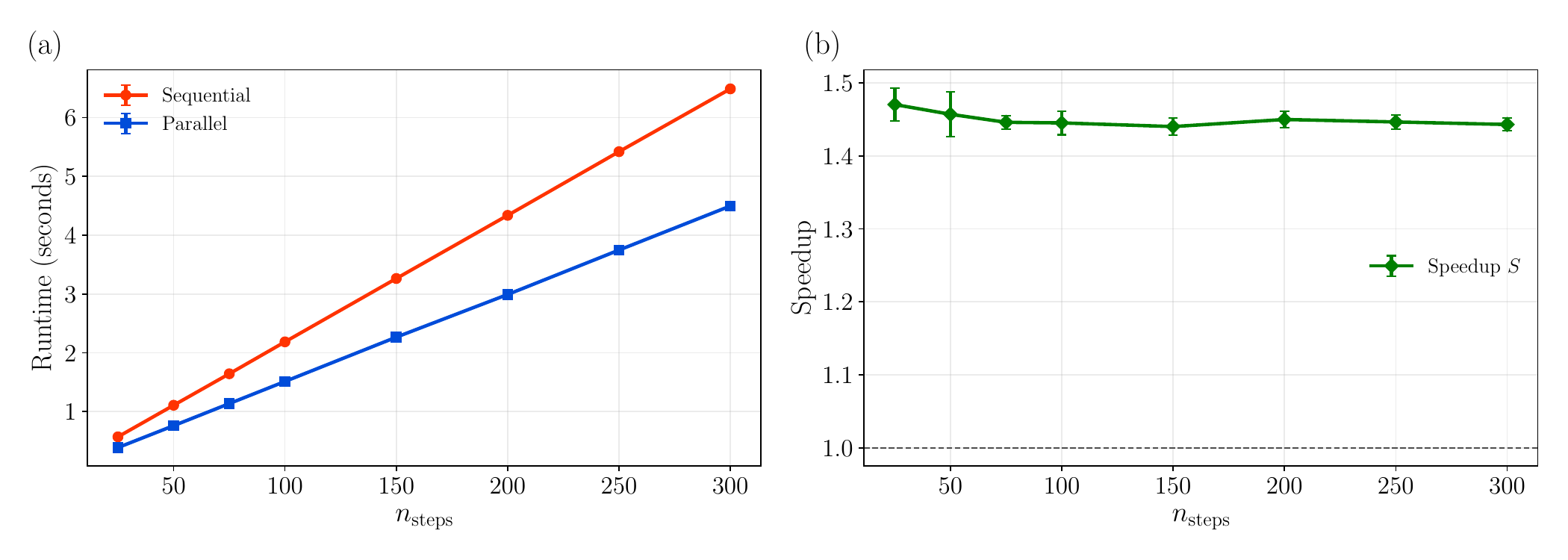}
 \caption{(a) Runtime of state-vector PITE simulations using parallel and sequential evolution as a function of the number of steps $n_{\rm steps}$, and (b) the corresponding speedup, $S = t_{\rm seq} / t_{\rm par}$. The test was performed for a Heisenberg chain with $L = 16$ sites. Error bars show the standard deviation over 10 simulations per data point.}
 \label{fig:parallel_speedup1}
\end{figure}

For smaller systems, however, the overhead associated with process synchronisation after each step outweighs the benefit of parallelisation, as shown in Fig.~\ref{fig:parallel_speedup2}. The speedup approaches $\sim 1.6 \times$ for large systems and does not depend on the number of PITE steps, since both evolution branches are evaluated independently at each step (see Fig.~\ref{fig:parallel_speedup1}).

Our benchmarks were performed on a MacBook Pro equipped with an Apple M1 Max processor. The exact speedup and the crossover system size depend on the specific hardware and software environment.

\subsection{Optimised state-vector evolution}

\begin{figure}[H]
 \centering
 \includegraphics[width=0.95\linewidth]{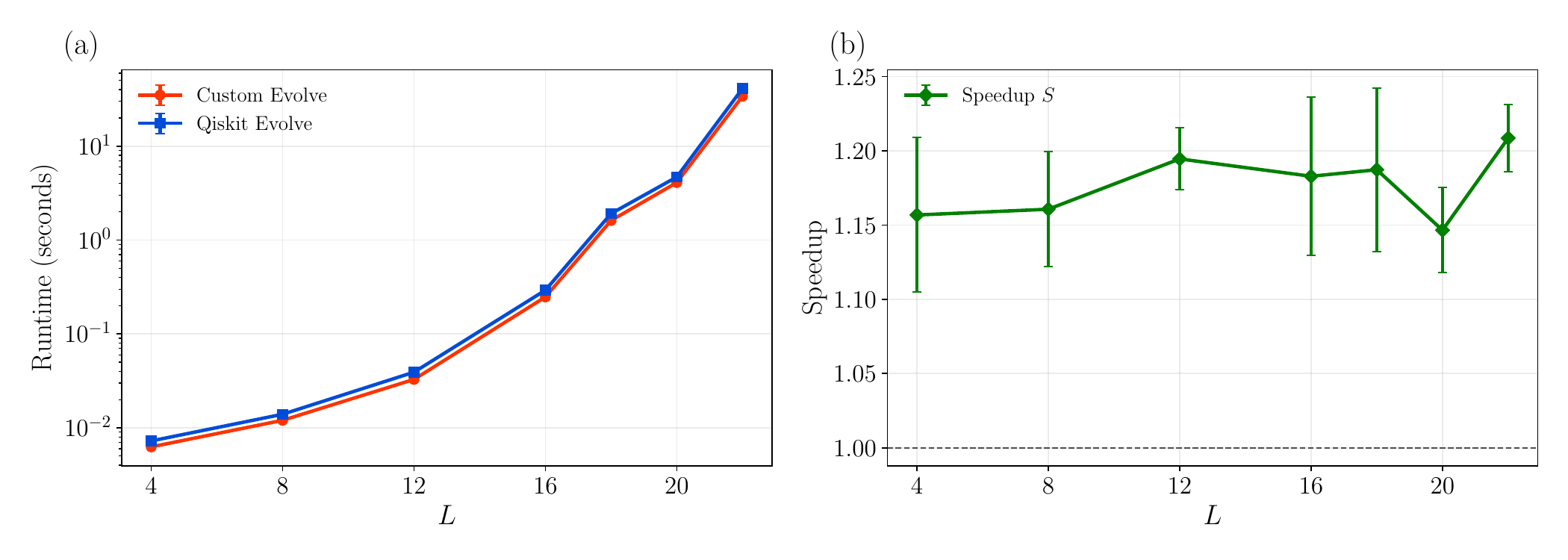}
 \caption{(a) Runtime of state-vector PITE simulations using the custom evolution routine and the Qiskit \texttt{evolve()} method as a function of the number of PITE steps $n_{\rm steps}$, and (b) the corresponding speedup. The test was performed for a Heisenberg chain with $L = 16$ sites. Error bars show the standard deviation over 10 simulations per data point.}
 \label{fig:evolve_speedup1}
\end{figure}

\begin{figure}[H]
 \centering
 \includegraphics[width=0.95\linewidth]{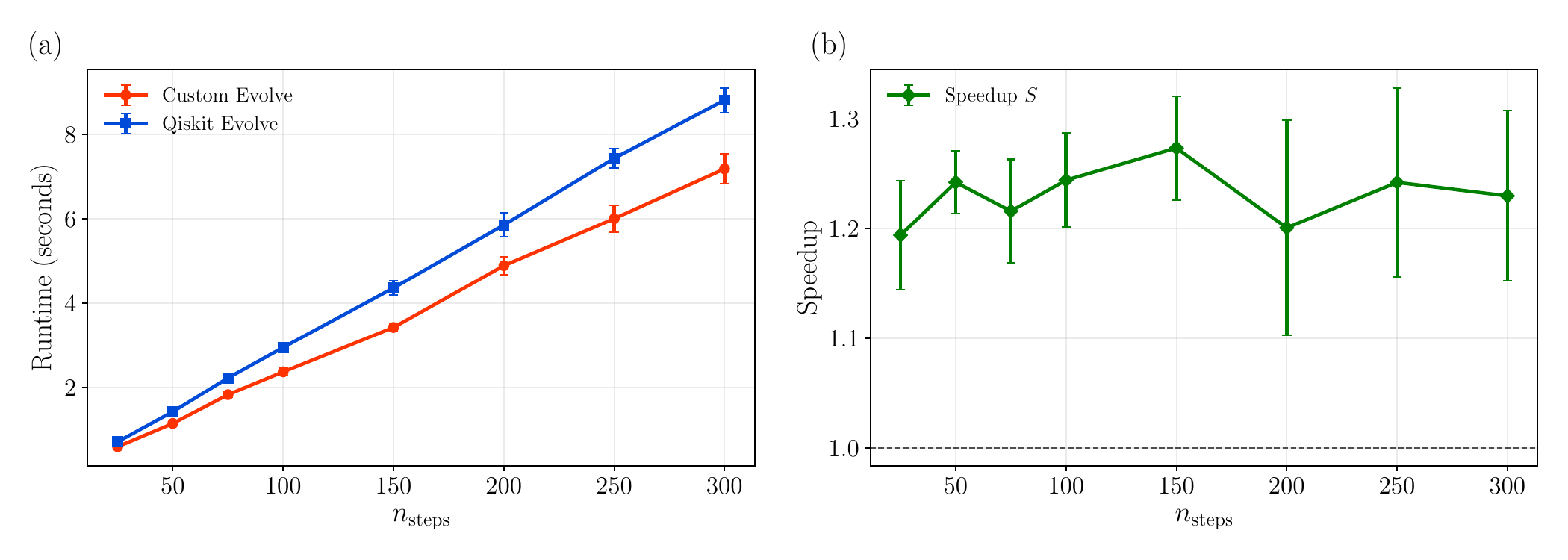}
 \caption{(a) Runtime of state-vector PITE simulations using the custom evolution routine and the Qiskit \texttt{evolve()} method as a function of the number of steps $n_{\rm steps}$, and (b) the corresponding speedup. The test was performed for a Heisenberg chain with $L = 16$ sites. Error bars show the standard deviation over 10 simulations per data point.}
 \label{fig:evolve_speedup2}
\end{figure}

Since the time-evolution circuit used in \SRC{} is always constructed from a \texttt{SparsePauliOp}, the standard \texttt{evolve()} function provided by Qiskit can be specialised to this case. By eliminating unnecessary \texttt{reshape()} operations, we achieve an additional speedup of $\sim 1.2 \times$, independent of system size and the number of time steps (see Figs.~\ref{fig:evolve_speedup1} and \ref{fig:evolve_speedup2}). 

The use of the custom optimised evolution routine is enabled by default and can be disabled by passing \texttt{use\_qiskit\_evolve=True} to the \texttt{run()} function.

Our \texttt{evolve\_statevector()} function can also be used independently of the PITE algorithm to perform real-time evolution, as demonstrated in Sec.~\ref{subsec:time_evo}.

\subsection{Computational cost of shot-based simulations}
\label{appsec:shot_performance}
\begin{figure}[H]
 \centering
 \includegraphics[width=0.95\linewidth]{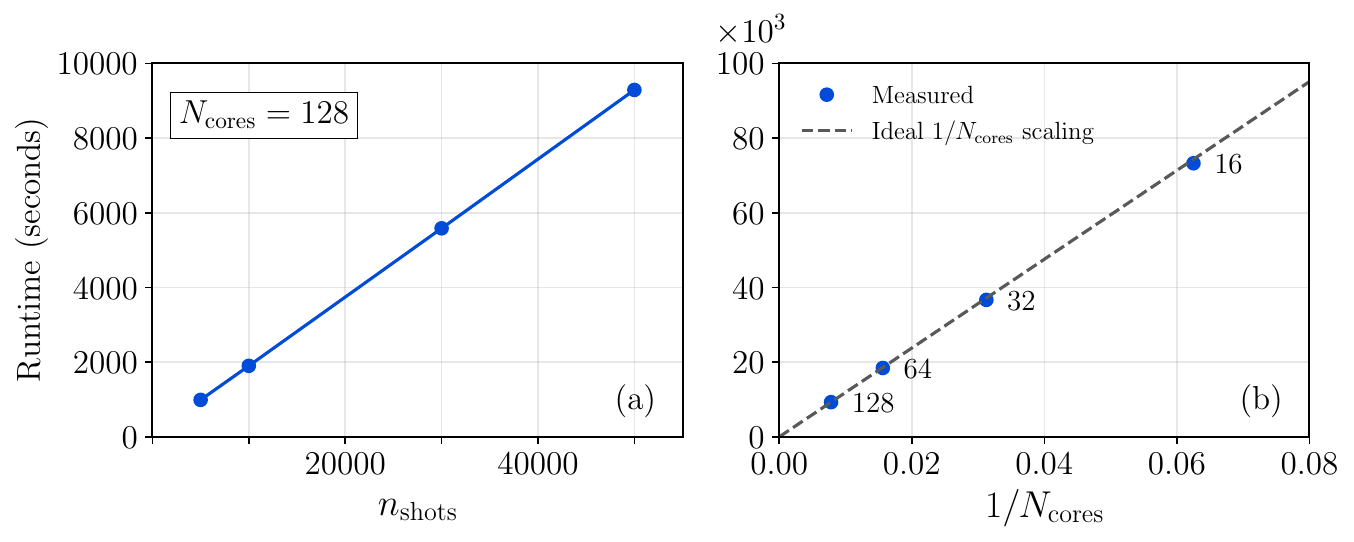}
 \caption{
Runtime of shot-based PITE simulations for the $L = 16$ Heisenberg chain.
(a) Dependence on the number of shots $n_{\mathrm{shots}}$ at fixed $n_{\mathrm{steps}} = 80$ using 128 CPU cores.
(b) Dependence on $1/N_{\mathrm{cores}}$ at fixed $n_{\mathrm{shots}} = 50{,}000$ and $n_{\mathrm{steps}} = 80$.
The dashed line indicates ideal $1/N_{\mathrm{cores}}$ scaling normalized to the largest-core measurement.
}
 \label{fig:shot_runtime}
\end{figure}

While the state-vector PITE algorithm for an $L=16$ Heisenberg chain can be executed in seconds on a laptop (see Fig.~\ref{fig:evolve_speedup2}), a shot-based simulation requires several hours on a 128-core workstation.
Figure~\ref{fig:shot_runtime} shows the corresponding runtimes.
Panel (a) shows that the runtime increases approximately linearly with the number of shots $n_{\mathrm{shots}}$, as expected since each shot requires an independent sampling run.
As shown in Fig.~\ref{fig:Heisenberg_Shot_vs_SV}, approximately $50{,}000$ shots are required to obtain a statistically reliable estimate. Using all 128 CPU cores of a single CARO node at DLR, this corresponds to a total runtime of $\sim 2.5$ hours for 80 steps. The node is equipped with two AMD EPYC 7702 processors with 64 cores each.
Panel (b) shows the parallel scaling with the number of CPU cores.
The runtime is plotted as a function of $1/N_{\mathrm{cores}}$, so ideal parallelization corresponds to a straight line through the origin.
The measured data closely follow the ideal $1/N_{\mathrm{cores}}$ scaling, indicating that the shot-based simulations are efficiently parallelized over the sampled trajectories in this regime.
Nevertheless, even with this favorable parallel scaling, the absolute computational cost remains substantial.

Moreover, each energy data point requires an independent simulation, since the energy is only measured after the final step. Consequently, obtaining the full energy convergence as a function of the number of steps $n_{\mathrm{steps}}$ leads to an overall computational cost that scales quadratically in $n_{\mathrm{steps}}$ (compared to the linear scaling of the state-vector simulation).
Shot-based simulations are therefore significantly more computationally demanding, making the state-vector formulation of PITE indispensable for parameter exploration.

\pagebreak
\section{Full Code Examples}
\label{appsec:code}
\renewcommand{\lstlistingname}{\SRC \textit{ Example Code}}

This appendix provides complete Python scripts for state-vector-based simulations, shot-based simulations, and ED simulations, as well as the full code corresponding to the basic workflow example in Sec.~\ref{sec:basic}. In addition, we present different ways of constructing 2D Hamiltonians via the \texttt{PauliStringOperator} interface. All scripts shown here, as well as additional examples, including the scripts used to generate the plots in this paper, are also included in the \texttt{examples} directory of the package available on \href{https://gitlab.com/dlr-sc-qc/many-body/svpite}{GitLab}.

\subsection{State-vector simulation}

\begin{lstlisting}[style=python, caption={State-vector-based simulation of the 1D four-site TFIM.}]
# examples/ising_4_sv.py
from svpite import PITEStatevector
from svpite.configs import PITEConfig
from svpite.operator.models import IsingHamiltonian

import matplotlib.pyplot as plt

H = IsingHamiltonian(n_sites=4, J=-1.0, h=-1.0, bc="PBC")
config = PITEConfig(gamma=0.78, n_steps=100, dt=0.1, order=1,
                    initial_state="plus")

algorithm = PITEStatevector(config, H)
result = algorithm.run(verbose=True, return_statevector=True, use_qiskit_evolve=False, parallel_evolve=True)

# Print formatted result summary
print(result)

plt.plot(range(config.n_steps+1), result.energies, marker='.')
plt.title("Ising 4-site Hamiltonian Energy vs PITE Steps")
plt.xlabel("PITE Steps")
plt.ylabel("$E_0/L$")
plt.show()

\end{lstlisting}

\subsection{Shot-based simulation}
\begin{lstlisting}[style=python, caption={Shot-based simulation of the 1D four-site TFIM.}]
# examples/ising_4_shot.py
from svpite import PITEShot
from svpite.configs import PITEConfig
from svpite.operator.models import IsingHamiltonian

import matplotlib.pyplot as plt

H = IsingHamiltonian(n_sites=4, J=-1.0, h=-1.0, bc="PBC")
config = PITEConfig(gamma=0.78, n_steps=40, dt=0.1, order=1,
                    n_shots=10000, initial_state="plus")

algorithm = PITEShot(config, H)
result = algorithm.run(verbose=True, return_circuit=True)

# Print formatted result summary
print(result)

# Draw and show the first-iteration PITE circuit
result.circuit.decompose(reps=3).draw(output="mpl")
plt.show(block=False)  # Show circuit without blocking

# Plot the probability
plt.figure()
plt.plot(result.probabilities[1:], marker='.')
plt.axhline(y=1.0, color='r', linestyle='--')
plt.text(0.2, 0.1, f"Energy per Site: {result.energy:.6f}\n #Shots: {config.n_shots}", transform=plt.gca().transAxes)
plt.xlabel("PITE step")
plt.ylabel(r"$\mathbb{P}_0$")
plt.show()

\end{lstlisting}

\subsection{ED simulation}

\begin{lstlisting}[
  style=python,
  caption={Exact diagonalisation of the 1D four-site TFIM.},
  label={lst:tfim-ed-quspin}
]
# examples/ising_4_ed.py
from svpite import ED
from svpite.configs import EDConfig
from svpite.operator.models import IsingHamiltonian

H = IsingHamiltonian(n_sites=4, J=-1.0, h=-1.0, bc="PBC")
spin_basis_1d_kwargs = {"kblock": 0, "pblock": 1}
# All options have default values. Here we explicitly set them as an example.
config = EDConfig(spin_basis_1d_kwargs=spin_basis_1d_kwargs,
                    tol=1e-10, maxiter=1000, v0=None)

algorithm = ED(config, H)
result = algorithm.run(verbose=True, return_groundstate=True)

# Print formatted result summary
print(result)

\end{lstlisting}

\subsection{2D Hamiltonian construction}
\label{appsubsec:2D_Hamiltonian_examples}
\begin{lstlisting}[style=python, caption={Different ways of constructing equivalent 2D Hamiltonians.}]
# examples/hamiltonian_construction_2d.py
from svpite.operator import Hamiltonian, models

J = -1.0
h = -1.0
shape = (2, 3)
bc = "OBC"

# 1. Predefined 2D Ising model
H1 = models.IsingHamiltonian(J=J, h=h, bc=bc, shape=shape)

# 2. Generic Hamiltonian with shape-aware uniform terms
H2 = Hamiltonian(shape=shape, terms_dict={"ZZ": (J, bc), "X": h})

# 3. Generic Hamiltonian from explicit 2D coordinate terms
zz_terms = []
x_terms = []
n_rows, n_cols = shape
for row in range(n_rows):
    for col in range(n_cols):
        x_terms.append([h, (row, col)])
        if col < n_cols - 1:
            zz_terms.append([J, ((row, col), (row, col + 1))])
        if row < n_rows - 1:
            zz_terms.append([J, ((row, col), (row + 1, col))])
H3 = Hamiltonian(shape=shape, terms_dict={"ZZ": zz_terms, "X": x_terms})

# 4. Manual add_term construction with 2D coordinate indices
H4 = Hamiltonian(shape=shape)
for coeff, sites in zz_terms:
    H4.add_term("ZZ", coeff, sites)
for coeff, site in x_terms:
    H4.add_term("X", coeff, site)

# 5. Build separate 2D Hamiltonian pieces and add them
H_zz = Hamiltonian(shape=shape, terms_dict={"ZZ": (J, bc)})
H_x = Hamiltonian(shape=shape, terms_dict={"X": h})
H5 = H_zz + H_x

# 6. Working with flattened indices
H6 = Hamiltonian(n_sites=shape[0] * shape[1])
for row in range(shape[0]):
    for col in range(shape[1]):
        flat_index = row * shape[1] + col
        H6.add_term("X", h, (flat_index,))
        if col < shape[1] - 1:
            H6.add_term("ZZ", J, (flat_index, flat_index + 1))
        if row < shape[0] - 1:
            H6.add_term("ZZ", J, (flat_index, flat_index + shape[1]))

for i, H in enumerate([H1, H2, H3, H4, H5, H6], start=1):
    print(f"H{i}: ", H)
print("H1 == H2 == H3 == H4 == H5 == H6: ", H1 == H2 == H3 == H4 == H5 == H6)

\end{lstlisting}

\pagebreak

\subsection{Basic workflow example}
\begin{lstlisting}[style=python, caption={Comparison of state-vector PITE, Shot-based PITE and ED}]
# examples/basic_workflow.py
# ---- Importing everything necessary ----
# Algorithms
from svpite import PITEStatevector, PITEShot, ED, evolve_statevector

# Configurations
from svpite.configs import PITEConfig, EDConfig

# Specific models
from svpite.operator.models import XXZHamiltonian

import numpy as np
    
# ---- Define model parameters ----
L = 8 # number of lattice sites
J = 1.0 / 4.0 # coupling constant
Delta = 1.0 / np.sqrt(2.0) # anisotropy
hz = 1.0 / 5.0 # external field
bc = 'PBC' # periodic boundary conditions   


# ---- Define XXZ Hamiltonian in external z-field ----
# Use standard XXZ model as starting point
H = XXZHamiltonian(L, J, Delta, bc)

# Add external z-field terms to the Hamiltonian
H.add_uniform_terms("Z", hz)


# ---- Define algorithm parameters ----
config = PITEConfig(
    gamma=0.72, # Gamma parameter of the PITE algorithm
    n_steps=80, # Number of imaginary-time evolution steps
    dt=0.2, # Imaginary-time step size
    order=1, # Trotter order for the real-time evolution gate
    reps=1, # Number of Suzuki-Trotter slices per PITE step
    initial_state='singlet', # Initial state of the evolution
    n_shots=10000, # Number of shots for shot-based PITE
)


# ---- Initialize and run algorithms ----
# Statevector PITE
sv_algorithm = PITEStatevector(config, H)
sv_result = sv_algorithm.run(verbose=True, return_statevector=True)

# Shot-based PITE
shot_algorithm = PITEShot(config, H)
shot_result = shot_algorithm.run(verbose=True, return_circuit=False)


# ---- Access and print results ----
print(sv_result)
print(shot_result)

# Statevector PITE results
E_sv = sv_result.energies  # Energy per site vs PITE step
P_sv = sv_result.probabilities  # Success probabilities vs PITE step

# Shot-based PITE results
E_shot = shot_result.energy  # Final energy per site
dE_shot = shot_result.energy_std  # Standard deviation of final energy per site
P_shot = shot_result.probabilities  # Success probabilities vs PITE step


# ---- Reference ED calculation ----
# Define ED configuration with symmetry sectors for the ground state computation
ed_basis_kwargs = {"Nup": L//2, "pblock": 1, "kblock": 0}
ed_config = EDConfig (
    tol = 1e-10, # Tolerance for eigensolver
    maxiter = 1e4, # Maximum iterations for eigensolver
    v0 = None, # Initial vector for eigensolver
    spin_basis_1d_kwargs = ed_basis_kwargs
)

# Initialize and run ED algorithm
ed_algorithm = ED(ed_config, H) 
ed_result = ed_algorithm.run(verbose=True, return_groundstate=False)

# Access ED results
print(ed_result)
E_ed = ed_result.energy # Computed ground state energy


# ---- Beyond the Basic workflow ----
# Get the final statevector from the results
psi0 = sv_result.final_statevector

# Real-time evolution generated by H
U_time = H.to_time_evolution_circuit(time=1, reps=100, order=1)

# Evolve the PITE output state
psi_t = evolve_statevector(psi0, U_time)

O = H.to_pauli_string_operator()
O_qiskit = O.to_sparse_pauli_op()
value_t = psi_t.expectation_value(O_qiskit)

# Access full statevector as NumPy array
psi_np = psi0.data

# Convert operator to dense matrix
O_mat = O.to_matrix(sparse=True)

# Example: expectation value using NumPy
value = np.vdot(psi_np, O_mat @ psi_np)

print("Real-time evolution successful!")


# ---- Plot results ----
import matplotlib.pyplot as plt
from matplotlib.ticker import MaxNLocator
plt.rcParams["text.usetex"] = True
plt.rcParams["text.latex.preamble"] = r"\usepackage{amsfonts}"
plt.rcParams["font.family"] = "serif"
plt.rcParams['font.size'] = 18
plt.rcParams['legend.fontsize'] = 16
plt.rcParams['xtick.labelsize'] = 16
plt.rcParams['ytick.labelsize'] = 16

cmap_red = plt.get_cmap("autumn")
cmap_blue = plt.get_cmap("winter")

fig, ax = plt.subplots(1, 2, figsize=(14, 5))

# Panel 1: PITE success probabilities
ax[0].plot(range(len(P_shot) - 1), P_shot[1:], 'o-', label='Shot-based PITE', color=cmap_red(0.2), markersize=3)
ax[0].plot(range(len(P_sv) - 1), P_sv[1:], 's-', label='Statevector PITE', color=cmap_blue(0.3), markersize=3)
ax[0].axhline(1.0, color='0.3', linestyle='--', linewidth=1)
ax[0].set_xlabel('$n_{\\rm steps}$')
ax[0].set_ylabel(r'$\mathbb{P}_0$')
ax[0].yaxis.set_major_locator(MaxNLocator(nbins=5))
ax[0].legend(frameon=False)
ax[0].grid(True, alpha=0.25)

# Panel 2: Energy convergence
ax[1].plot(range(len(E_sv)), E_sv, 's-', label='Statevector PITE', color=cmap_blue(0.3), markersize=3)
ax[1].axhline(E_ed, color='black', linestyle='--', label=f'ED Energy = {E_ed:.4f} ')
ax[1].axhline(E_shot, color=cmap_red(0.2), linestyle='-', label=f'Shot-based PITE = {E_shot:.4f} $\pm$ {dE_shot:.4f}')
ax[1].fill_between(range(-5, len(E_sv)+5), E_shot - dE_shot, E_shot + dE_shot, alpha=0.2, color=cmap_red(0.2))
ax[1].set_xlim(-5, len(E_sv)+4)
ax[1].set_xlabel('$n_{\\rm steps}$')
ax[1].set_ylabel(r'$E_0/L$')
ax[1].yaxis.set_major_locator(MaxNLocator(nbins=5))
ax[1].legend(frameon=False)
ax[1].grid(True, alpha=0.25)

for i, a in enumerate(ax):
    a.text(-0.08, 1.02, f'({chr(97+i)})',
        transform=a.transAxes,
        ha='left', va='bottom', fontsize=20)

fig.tight_layout()
fig.savefig("Basic_Workflow_Example_Panels.pdf", bbox_inches='tight', transparent=True)
plt.show()


\end{lstlisting}

% \section{About references}
% Your references should start with the comma-separated author list (initials + last name), the publication title in italics, the journal reference with volume in bold, start page number, publication year in parenthesis, completed by the DOI link (linking must be implemented before publication). If using BiBTeX, please use the style files provided  on \url{https://scipost.org/submissions/author_guidelines}. If you are using our LaTeX template, simply add
% \begin{verbatim}
% \bibliography{your_bibtex_file}
% \end{verbatim}
% at the end of your document. If you are not using our LaTeX template, please still use our bibstyle as
% \begin{verbatim}
% \end{verbatim}
% in order to simplify the production of your paper.
\end{appendix}

% \bibliographystyle{SciPost_bibstyle}
% \bibliography{ref.bib}

\end{document}